\newcommand{\beq}{\begin{eqnarray}}
\newcommand{\eeq}{\end{eqnarray}}
\newcommand{\be}{\begin{equation}}
\newcommand{\ee}{\end{equation}}

\def\la{\mathrel{\mathpalette\fun <}}

\def\fun#1#2{\lower3.6pt\vbox{\baselineskip0pt\lineskip.9pt
\ialign{$\mathsurround=0pt#1\hfil ##\hfil$\crcr#2\crcr\sim\crcr}}}

\newcommand{{\SD}}{\rm SD}
\newcommand{\pp}{\prime\prime}

\newcommand{\vex}{\mbox{\boldmath${\rm x}$}}

\newcommand{\ver}{\mbox{\boldmath${\rm r}$}}
\newcommand{\vesig}{\mbox{\boldmath${\rm \sigma}$}}

\newcommand{\vep}{\mbox{\boldmath${\rm p}$}}
\newcommand{\veq}{\mbox{\boldmath${\rm q}$}}

\newcommand{\veQ}{\mbox{\boldmath${\rm Q}$}}

\newcommand{\vek}{\mbox{\boldmath${\rm k}$}}
\newcommand{\ven}{\mbox{\boldmath${\rm n}$}}
\newcommand{\veu}{\mbox{\boldmath${\rm u}$}}
\newcommand{\vev}{\mbox{\boldmath${\rm v}$}}

\newcommand{\veal}{\mbox{\boldmath${\rm \alpha}$}}

\newcommand{\lan}{\langle}
\newcommand{\ran}{\rangle}

\newcommand{\tr}{{\rm tr} \,}

\documentclass[showpacs,nofootinbib]{revtex4}%prc,preprint,nofootinbib

\usepackage{hyperref}         %Гипертекстовые ссылки
\usepackage{epsfig,graphicx}
\usepackage{amssymb}
\usepackage{amsfonts}
\usepackage{bm}
\usepackage{graphics}
\usepackage{epsfig}
\usepackage{mciteplus}

\usepackage{epstopdf}

\def\inbar{\,\vrule height1.5ex width.4pt depth0pt}
\def\IR{\relax{\rm I\kern-.18em R}}
\def\IC{\relax\hbox{$\inbar\kern-.3em{\rm C}$}}

\begin{document}

\title{Channel coupling in heavy quarkonia: energy levels, mixing, widths and new states}

\author{\firstname{I.~V.}~\surname{Danilkin}}
\email{danilkin@itep.ru}\affiliation{Moscow Engineering Physics
Institute, Moscow, Russia} \affiliation{Institute of Theoretical
and Experimental Physics, Moscow, Russia}

\author{\firstname{Yu.~A.}~\surname{Simonov}}
\email{simonov@itep.ru} \affiliation{Institute of Theoretical and
Experimental Physics, Moscow, Russia}


\begin{abstract}
The  mechanism of channel coupling via decay products is used to
study energy shifts, level mixing as well as the possibility of
new near-threshold resonances in $ c\bar c, b\bar b$ systems. The
Weinberg eigenvalue method is formulated in the multichannel
problems, which allows to describe coupled-channel resonances and
wave functions in a unitary way, and to predict new states due to
channel coupling. Realistic wave functions for all single-channel
states and decay matrix elements computed earlier are exploited,
and no new fitting parameters are involved. Examples of level
shifts, widths and mixings are presented; the dynamical origin of
$X(3872)$ and the destiny of the single-channel $~2^3P_1(c\bar c)$
state are clarified. As a result a sharp and narrow peak  in the
 state with quantum numbers $J^{PC}=1^{++}$  is found
 at 3.872 GeV, while the single-channel resonance originally  around 3.940 GeV,
becomes increasingly broad  and disappears with growing  coupling to open
channels.

\end{abstract}

\maketitle

\section{INTRODUCTION}
\label{sect.1}

Most hadron states are coupled by strong interaction to closed or
open decay channels, and thus are subjects of the Theory of
Strongly Coupled  Channels (TSCC). The latter topic was developed
during many decades, see
\cite{Eichten:1979ms,*Eichten:1978tg,*Eichten:1975ag,Geiger:1992va,
*Geiger:1991qe,*Geiger:1991ab,*Geiger:1989yc,vanBeveren:1982qb,*vanBeveren:1979bd,Tornqvist:1995kr,*Tornqvist:1995ay}
and \cite{Badalian:1981xj} for a review, and also
\cite{Kalashnikova:2005ui,*Baru:2003qq,Barnes:2007xu,Pennington:2007xr}
as more recent publications. In the present paper we  apply TSCC
specifically to  the case of Okubo-Zweig-Iizuka rule allowed
two-body decay channels of charmonia and bottomonia. In doing so
we need several prerequisites. First of all, it is the one-channel
description of charmonia and bottomonia as $c\bar c$, $b\bar b$
states in relativistic Hamiltonian formalism
\cite{Dubin:1993fk,*Badalian:2008sv}, developed in the framework
of the Field Correlator method (FCM)
\cite{Dosch:1987sk,*Dosch:1988ha,*Simonov:1987rn} (see
\cite{DiGiacomo:2000va} for a review) and    wave functions of
stable states in $x$ or $p$ space cast in the numerical form. The
latter have been accurately computed using this method with only
universal input: the string tension $\sigma$, the current (pole)
quark masses $m_i$, and the strong coupling $\alpha_s(q)$
\cite{Badalian:2008bi,*Badalian:2004gw,*Badalian:2008dv,*Badalian:2007zz}.

The next ingredient is an effective relativistic Lagrangian for
the pair creation, inducing the string breaking. To this end we
are using  the decay mass vertex $\int \bar{\psi}M_\omega\psi\,
d^{\,4} x$ introduced in \cite{Simonov:2007bm} and exploited for
dipion transitions in
\cite{Simonov:2007bm,Simonov:2008qy,Simonov:2008ci} and for the
reaction channel $\Upsilon (nS) \rightarrow B\bar B, B\bar{B} \pi$
in \cite{Simonov:2008cr}. In principle, $M_\omega$  can be
expressed  in terms of quark masses and average energies, but we
use it as  the only one parameter, which is fixed in our previous
studies
\cite{Simonov:2007bm,Simonov:2008qy,Simonov:2008ci,Simonov:2008cr}
Finally, as it was shown in \cite{Simonov:2007bm} and before in
\cite{Barnes:2005pb,*Barnes:1991em,Ackleh:1996yt} the transition
matrix element reduces to the overlap integral of wave functions
of decaying system and products of decay. It is interesting, that
the vertex operator in this integral contains not only $M_\omega$,
but also the $Z$ factor of the decay process constructed from the
Dirac trace of all involved hadron vertex states, and projection
operators. This technic,  introduced in \cite{Simonov:2007bm}, is
a relativistic equivalent of the nonrelativistic one with
spin-angular momentum (Clebsch-Gordon) coefficients used in the
framework of $^3P_0$ model
\cite{Micu:1968mk,*LeYaouanc:1972ae,*LeYaouanc:1977ux,*LeYaouanc:1977gm}.
As a result one obtains a system of integrodifferential equations
for new wave functions and energy eigenvalues, which can be easily
solved in the lowest approximation for energy shifts, widths and
level mixing coefficients. At the same time we have developed a
$(2\times 2)$ variant of wave functions and matrix elements for
light quarks in heavy-light mesons. Several examples of this kind
are shown below.

At a deeper level, one meets with several problems: (i) First, the
states above decay  thresholds  are unstable and the definition of
the wave function itself is questionable in a rigorous sense,
since an admixture of continuous  spectrum states appears. Here
different approaches exist. The most rigorous is the Weinberg
procedure \cite{Weinberg:1963zz,*newton:319,*Smithies1958}, named
below the Weinberg Eigenvalue Method (WEM). It is used to define
the resonance wave function and energy, as well as $t$-matrix via
Weinberg eigenvalues. The WEM has been used before in the
one-channel 2-body and 3-body problems
\cite{Weinberg:1963zz,*newton:319,*Smithies1958},\cite{Narodetsky:1980nx,*Narodetsky:1969jf,*Herzenberg,*Fuda}.
We have found that it is specifically useful in case of TSCC,
since coupled channels (CC) induce  the energy-dependent force
term, which violates standard orthonormalization procedure, while
this term can easily be treated in WEM. (ii) Second, even the
closed channels cause the problems. In terms of hadron loops it
was treated in many papers, see e.g.
\cite{Geiger:1992va,*Geiger:1991qe,*Geiger:1991ab,*Geiger:1989yc}
and recent papers \cite{Barnes:2007xu,Pennington:2007xr}, where
some theorems were formulated \cite{Barnes:2007xu} and the
renormalization method was suggested \cite{Pennington:2007xr}.

In essence the problem here is similar to the problem of
unquenched quark pairs. It occurs also for stable hadrons, where
the renormalozation procedure is  necessary in general. We will
not discuss these topics in the given paper, assuming that the
renormalization is done e.g., by readjusting the pole quark mass.
We also disregard the important topic of full relativistic
invariance for composite objects moving with different velocities,
e.g. charmonium decaying in its c.m. system into heavy-light
mesons, with their wave functions defined in their c.m. systems.
This is done assuming small relative velocities near thresholds.
Far from thresholds these factors become important. Similarly,
near thresholds we assume here, as well as in
\cite{Simonov:2007bm,Simonov:2008qy,Simonov:2008ci,Simonov:2008cr},
the $^3P_0$ type of the decay vertex, while at higher energies
this type of decay may be replaced by another one, e.g. the
$^3S_1$ type.

In this paper we systematically apply WEM to find the shifts and
widths of $(n^3S_1)$ energy levels, as well as mixing between
them. We find the method to be especially useful to discover the
analytic structure and pole positions in the case of strong CC. A
particular example of the  $2^{\,3}P_1$ level proves to be a good
illustration of our analysis. From the experimental point  of view
two interesting problems appear. First one, why in experiment only
the peak at the lowest $D_0D^*_0$ threshold is seen, while at the
slightly higher, $D_+D_-^*$ no peak was ever seen? Secondly,
possible resonances at 3.940 GeV found in
\cite{Abe:2007jn,*Pakhlova:2008di,*Yuan:2009iu}, seemingly are not
$J=1$,  and very likely the $1^{++}$ state around 3.940 GeV was
never observed. Our analysis allows  to  answer both questions at
the same time, as will be explained below.

As a result we find two poles due to a single eigenvalue in the possitions near
3.872 GeV and 3.940 GeV, but the latter peak becomes too wide  and  finally
disappears with increasing $CC$, which can explain the experimental situation
\cite{Abe:2007jn,*Pakhlova:2008di,*Yuan:2009iu}. Moreover, retaining  the peak
appears at the lower threshold 3.872 GeV, and not at the higher threshold 3.879
GeV, again in agreement with experiment
\cite{Abe:2007jn,*Pakhlova:2008di,*Yuan:2009iu}.

The plan of the paper is as follows. In the next section we introduce the
general formalism for the Green's functions of charmonia and bottomonia with
the inclusion of decay channels. We present equations for wave functions
(Green's functions) both in $Q\bar Q$ and $(Q\bar q) (\bar Q q)$ channels. In
section \ref{sect.3} the CC resonances are considered in the decay channel and
condition for the existence of a CC resonance is formulated. In section
\ref{sect.4} the rigorous Weinberg theory of CC resonances is presented. In
section \ref{sect.5} the mixing of states in WEM is considered. In section
\ref{sect.8} we present results for values of level shifts and widths for
$3^{\,3}S_1$ state and also mixing between $3^{\,3}S_1$ and $2^{\,3}S_1$ state,
as well as the analysis of the situation  in the $2^{\,3}P_1$ state. In section
\ref{sect.9} summary and prospectives are given.
%In section \ref{sect.6} the analytic structures of WEM amplitudes
%and pole positions are derived.

\section{GENERAL FORMALISM OF STRING-BREAKING CHANNEL COUPLING}
\label{sect.2}

We consider two sectors of hidden  and open flavor with initial
and final bare gauge invariant operators, for heavy quarkonium
sector,

\textbf{I.} $ j^{\,(I)}_i (x) = \bar \psi_Q(x)\, \Gamma_i\, \psi_Q
(x)$
\\
and for heavy-light meson sector,

\textbf{II.} $j^{\,(II)}_i(x) =\bar\psi_Q (x)\, \Gamma_i\, \psi_q
(x)$,\\
where  $\Gamma_i=1,\,\gamma_\mu,\,...,  D_\mu
\sigma_{\mu\nu},....$ With the help of $j_i^{\,(I,II)}$ one
generates bare mesons and as shown in
\cite{Simonov:2007bm,Badalian:2007km} one can project physical
amplitudes \footnote{Note, that the procedure of hadron state
projection is here fully equivalent to that used in lattice
approach, see e.g.
\cite{Hashimoto:1999yp,*Becirevic:2004ya,*Brommel:2007wn,*Bowler:1996ws}.}
(Green's functions) with physical wave functions
$\Psi^{(n_1)}_{Q\bar Q},\, \psi^{(n_2)}_{Q\bar q},\,
\psi^{(n_3)}_{\bar Q q}$. For stationary states one can use
Green's functions in energy representation, e.g.
\begin{equation}
G^{(0)}_{Q\bar Q} \left(1,2;\,E\right) =\,\sum_{n_1}
\frac{\Psi^{(n_1)}_{Q\bar{Q}} (1)\,
\Psi^{\dag(n_1)}_{Q\bar{Q}}(2)}{E_{n_1}-E}=\,\frac{1}{H_0-E}\,.\label{1}
\end{equation}
Here superscript (0) of Green's function refers to the bare case,
when sector II is switched off, and $\Psi^{(n_1)}_{Q\bar{Q}},
E_{n_1}$ refer to the eigenfunctions and eigenvalues of the
relativistic string Hamiltonian $H_0$
\cite{Dubin:1993fk,*Badalian:2008sv}, for charmonium those were
calculated in
\cite{Badalian:2008bi,*Badalian:2004gw,*Badalian:2008dv,*Badalian:2007zz}
and for bottomonium in
\cite{Badalian:2008ik,*Badalian:2009bu,*Badalian:2004xv}. In
sector II the counterpart of (\ref{1}) consists of Green's
function of the pair $(Q\bar q), (q\bar Q)$. We neglect in the
first approximation interaction of two color singlet mesons, and
write the c.m. Green's function as
\begin{equation}
G^{(0)}_{Qq\,\bar q\bar Q} \left(1\bar 1 |2\bar 2;\, E\right)
=\,\sum_{n_2, n_3} \frac{\Psi_{n_2\,n_3} (1,\bar 1)\,
\Psi_{n_2\,n_3}^\dag(2,\bar 2)}{E_{n_2n_3}
(\vep)-E}\,\,d\Gamma(\vep)\,.\label{2}
\end{equation}

At this point we must take into account  a possible transition
(decay) of states in sector I into states of sector II, which can
be done in several ways. In literature it is common to assume one
of several types of phenomenological decay Lagrangians, e.g. $^3
P_0$ type
\cite{Micu:1968mk,*LeYaouanc:1972ae,*LeYaouanc:1977ux,*LeYaouanc:1977gm},
with vector confinement vertex
\cite{Eichten:1979ms,*Eichten:1978tg,*Eichten:1975ag}, or with
scalar confinement vertex, studied in \cite{Ackleh:1996yt}. For
bottomonium  a relativistic string decay vertex of  the following
form was used in
\cite{Simonov:2007bm,Simonov:2008qy,Simonov:2008ci,Simonov:2008cr}
\begin{equation}
\mathcal{L}_{sd}= \int \bar \psi_q\, M_\omega\, \psi_q\, d^{\,4} x
,~~ M_\omega =const \label{3}
\end{equation}
where $M_\omega$ was taken to be constant, $M_\omega \approx 0.8$
GeV from decays of bottomonium into $B\bar B, B\bar B^*, ...$
\cite{Simonov:2008ci,Simonov:2008cr}.

It is important, that we work in the c.m. system and consider both
wave functions and Hamiltonian obtained in the instantaneous
hyperplane, when all time coordinates of all particles are the
same\footnote{We omit boost corrections here, which makes
application of our method justifiable only close to thresholds.
Far from thresholds one should take into account both  boost and,
more important, a possible change in the pair creation vertex,
since high energy transfer to the $q\bar{q}$ pair might require
gluon exchange mechanism, hence $~^3S_1$ vertex}. Therefore the
vertex $\mathcal{L}_{sd}$ enters between instantaneous wave
functions of $Q\bar Q$ on one side and product of $Q\bar q$,
$q\bar Q$ on another side
\begin{equation}
J_{123} \equiv \frac{1}{\sqrt{N_c}}\int\bar y_{123} \Psi^+_{Q\bar
Q}\, M_\omega\, \psi_{Q\bar q}\, \psi_{q\bar  Q} \,d\tau.
\label{5}
\end{equation}

At this point one should define exactly how spin, momentum and coordinate
degrees of freedom enter in (\ref{5}), which we denote by  an additional factor
$\bar y_{123}$ and yet undefined phase space factor $d\tau$. One way  is to
exploit nonrelativistic-type decomposition which is  used in $^3P_0$
calculations
\cite{Micu:1968mk,*LeYaouanc:1972ae,*LeYaouanc:1977ux,*LeYaouanc:1977gm}. In
our considerations we are using two different ways. Below we begin from the
fully relativistic formalism of Dirac traces and projection operators, started
in \cite{Badalian:2007km} for decay constants and in \cite{Simonov:2007bm} for
dipion transitions and then will go into reduced $(2 \times 2)$ spin-tensor
formalism, explained in Appendix B. Note that the relativistic formalism with
$Z$ factors is similar  in lattice calculations for transition matrix elements,
see e.g.
\cite{Hashimoto:1999yp,*Becirevic:2004ya,*Brommel:2007wn,*Bowler:1996ws}.

\begin{figure}[t]
\begin{center}
  \includegraphics[angle=0,width=0.3\textwidth]{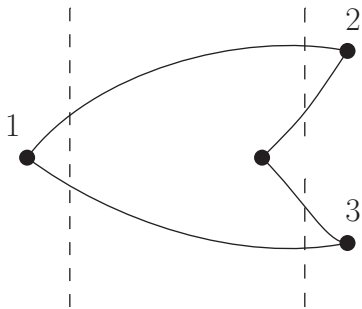}
  \caption{Decay matrix vertices.}
  \label{figure:Decay matrix vertices}
\end{center}
\end{figure}

In the formalism one considers initial and final meson creation
operators I, II, given above (see  in Appendix B   the Table
\ref{tab.5} of lowest operators and their $(2\times2)$ forms), and
composes the decay matrix element as shown in Figure 1 where
vertices $1,2,3,x$ are operators $\Gamma_{i=1,2,3,x}$ entering in
the bilinears $j_{\,i}= \bar \psi\, \Gamma_i\, \psi$, and lines
$(1,2), (2,x),...$ denote quark Green's functions $S_Q(1,2),\,
S_{\bar q} (2,x)$, etc. so that matrix element corresponding to
Figure \ref{figure:Decay matrix vertices} is
\begin{equation}
 S(1,2,x,3) = \,{\tr }
\,\Big\{\Gamma_1\, S_Q (1,2)\, \Gamma_2\, S_{\bar
q}(2,x)\,\Gamma_x\, S_q (x,3)\, \Gamma_3\, S_{\bar Q
}(3,1)\Big\}.\label{6}
\end{equation}
As was shown in \cite{Badalian:2007km,Simonov:2007bm} in the
approximation where one neglects influence of spin forces on wave
functions, one can replace
\begin{equation}
 S_{Q,q}=\frac{\left(m_{Q,q}+\omega_{Q,q}
\gamma_4-ip_i^{Q,q} \gamma_i\right)}{2\,\omega_{Q,q}}\,
=\Lambda^+_{Q,q} G_{Q,q},~~ S_{\bar Q\bar q}
=\frac{\left(m_{Q,q}-\omega_{Q,q} \gamma_4+ip_i^{Q,q}
\gamma_i\right)}{2\,\omega_{Q,q}}\, G_{\bar Q,\bar
q}=\Lambda^-_{Q,q} G_{\bar Q,\bar q}.\label{7}
\end{equation}
where $G_{Q,q}$ is quadratic  Green's function,
$\Lambda^\pm_{Q,q}$ are projection operators, the variables
$\omega_{Q,q}$ are the averaged kinetic energies and $m_{Q,q}$ are
the pole masses.

Our physical matrix element corresponding to the decay
$\Psi_{Q\bar Q} ^{(n_1)} \to \,\psi_{Q\bar q}^{(n_2)}
\psi^{(n_3)}_{\bar Q q}$ can be obtained from (\ref{6}) by
projecting the chosen intermediate states, as shown by dashed
lines in Figure \ref{figure:Decay matrix vertices}.

As a result as shown in \cite{Simonov:2007bm,Simonov:2008qy} the
physical projected matrix element has the form
%(in the relativistic formalism we write here $\bar y_{123}^{Rel}$)
%
\begin{equation}
J_{n_1n_2n_3}(\vep) = \frac{M_\omega}{\sqrt{ N_c}} \int \bar
y_{123}^{Rel}\, \Psi^{(n_1)}_{Q\bar Q} (\veu-\vev)\,
e^{i\,\vep\ver} \psi^{(n_2)}_{Q\bar{q}} (\veu-\vex)\,
\psi^{(n_3)}_{\bar{Q}q} (\vex-\vev)\, d^{\,3} \vex\, d^{\,3}
(\veu-\vev)\,,\label{8}
\end{equation}
where $N_c$ is the number of colours, $\ver =c\,(\veu-\vev),~~ c=
\frac{\omega_Q}{\omega_Q+\omega_q}$ and $\bar y_{123} = \frac{\bar
Z_{123}}{\sqrt{\prod^3_{i=1} \bar Z_i}}$. The expressions for
$\bar Z_{123}$, $\bar Z_i$ are proportional to Dirac traces of the
projector  operators and are given in [13,14]. We point that the
w.f
$\Psi^{(n_1)}_{Q\bar{Q}},\,\psi^{(n_2)}_{Q\bar{q}},\,\psi^{(n_3)}_{\bar{Q}q}$
in (\ref{8}) are no longer full w.f. of mesons,  but the radial
part
$R^{(n_1)}_{Q\bar{Q}},\,R^{(n_2)}_{Q\bar{q}},\,R^{(n_3)}_{\bar{Q}q}$
divided by $\sqrt{4\pi}$, while the angular part of the w.f. is
accounted  for in the factor $\bar{y}_{123}\,(\bar
y^{red}_{123})$. Important role is played by average values of
quark kinetic energies, $\omega_{Q,q} =\lan
\sqrt{m^2_{Q,q}+\vep^2}\,\ran$ inside heavy-light mesons in their
c.m. systems (if one neglects c.m. motion of these mesons). The
numbers of $\omega_Q,\, \omega_q$ are  computed from the
relativistic string Hamiltonian in
\cite{Badalian:2008bi,*Badalian:2004gw,*Badalian:2008dv,*Badalian:2007zz,Badalian:2008ik,*Badalian:2009bu,*Badalian:2004xv}.
In practical calculations it is more useful to exploit the
$(2\times2)$ reduction of the bispinor wave functions in
(\ref{5}), as $\psi =\left(\begin{array}{l}v\\w\end{array}\right)$
and  $\bar \psi\, \Gamma \,\psi = (v^c, w^c)\, \gamma_2 \gamma_4\,
\Gamma\left(\begin{array}{l}v\\w\end{array}\right)$, see details
in Appendix B. The resulting matrix element has the same form as
in (\ref{8}) but with $M_\omega\,\bar y_{123}\rightarrow
\gamma\,\bar y^{red}_{123}~$, where $\bar y^{red}_{123}$ is given
in Table \ref{tab.6} and $\gamma$ is proportional $M_\omega$ (see
Appendix C).

Now one can define the selfenergy part in sector I due to sector
II in the intermediate state, which is
\begin{equation}
w_{nm} (E) =\int \frac{d^{\,3} \vep}{(2\pi)^3} \sum_{n_2n_3}
\frac{J_{nn_2n_3} (\vep)\, J^+_{mn_2n_3}
(\vep)}{E-E_{n_2n_3}(\vep)}\label{9}
\end{equation}
and the total Green's function in sector I can  be written as (sum
over bound  states only)
\begin{equation}
 G^{(I)}_{Q\bar Q} (1,2;\,E) =\sum_n \frac{\Psi^{(n)}_{Q\bar Q}
(1)\, \Psi^{+(n)}_{Q\bar Q} (2)}{E_n-E}-\sum_{n,m}
\frac{\Psi^{(n)}_{Q\bar Q} (1)\, w_{nm} (E)\,  \Psi^{+(m)}_{Q\bar
Q} (2)}{(E_n-E)(E_m-E)}+...\label{10}
\end{equation}
where ellipsis implies terms of higher order in $w_{nm}$ and this
can be summed up as
\begin{equation}
G^{(I)}_{Q\bar Q} (1,2;E) =\sum_{n,m} {\Psi^{(n)}_{Q\bar Q}
(1)}{\,\big(\hat E-E+\hat w\big)^{-1}_{nm}\,}{\Psi^{+(m)}_{Q\bar
Q}} (2)\label{11}
\end{equation}
where matrix $(\hat E)_{nm} = E_n\,\delta_{nm}$.

Note, that in (\ref{11}) the Green's function is actually a
projection of  the coupled-channel system on the original
unperturbed $Q \bar Q$ wave functions $\Psi^{(n)}_{Q\bar Q}$. In
reality wave functions of the coupled-channel system  differ from
the latter and acquire continuous spectrum  pieces above the decay
threshold, and hence need a special treatment to be discussed
below.

The new spectrum is obtained from (\ref{11}) as
\begin{equation}
\det \big(E-\hat E-\hat w\big) =0\label{12}
\end{equation}
for one level in sector I it simplifies
\begin{equation}
E=E_n + w_{nn}
(E)\label{13}
\end{equation}
which yields energy shift and width in the first order
approximation in $\hat w$:
\begin{equation}
E^{\,(1)}_n = E_n + \textrm{Re}\,\big(w_{nn} (E_n)\big),~~
\Gamma_n^{(1)} = 2\, \textrm{Im} \,\big(w_{nn} (E_n)\big).
\label{14}
\end{equation}
In  the next order one should solve the transcendental in $E$
one-channel equation (\ref{13}), which is valid when $w_{nn}$ is
large, but $|w_{nm}| \ll |E_n-E_m|$.

Below the decay threshold one can diagonalize the matrix in
(\ref{11}) with unitary matrices
\begin{equation}
\big((E-\hat E -\hat w)^{-1}\big)_{nm} = U^+_{n\lambda} (E)
\frac{1}{E-E_\lambda}\, U_{\lambda m} (E)\label{15}
\end{equation}
and the Green's function acquires the form
\begin{equation}
 G_{Q\bar Q} ^{(I)} = \sum_\lambda \Phi_\lambda \frac{1}{E_\lambda-E}\,
\Phi^+_\lambda , ~~~ \Phi_\lambda = \sum_n \Psi^{(n)}_{Q\bar Q}\,
U^+_{n\lambda}(E).\label{16}
\end{equation}
In this way $\Phi_\eta$ become new orthogonal states comprising
all effects of mixture between bound states due to closed
channels. The same procedure can be applied for open channels
(above the decay threshold) when one neglects the widths of the
levels, i.e. imaginary part of $\hat w$.

One can define interaction $V_{121} $ in sector I due to sector
II,
\begin{equation}
V_{121} (\ver,\,\ver') = \sum_{n_2 n_3}
G^{(0)}_{n_2n_3}(\ver-\ver') \,X_{n_2 n_3}(\ver)\,
X^+_{n_2n_3}(\ver') \label{17}
\end{equation}
with
\begin{equation}
G^{(0)}_{n_2 n_3}(\ver - \ver') =\int\frac{d^{\,3}\vep}{(2\pi)^3}
\frac{e^{i\,\vep (\ver-\ver')}}{E_{n_2n_3} (\vep)-E}\,, \label{18}
\end{equation}
\begin{equation}
X_{n_2 n_3} (\ver) =\frac{M_\omega}{\sqrt{N_c}}\int \frac{d^{\,3}
\veq}{(2\pi)^3}\, e^{i\,\veq \ver}\, \psi^{(n_2)}_{Q\bar{q}}
(\veq)\, \psi^{(n_3)}_{\bar{Q}q} (\veq)\label{19}
\end{equation}
or, in momentum space
\begin{equation} V_{121} (\veq,\, \veq')=\sum_{n_2n_3} \int
\frac{d^{\,3}\vep}{(2\pi)^3}\,
\frac{X_{n_2n_3}(\veq-\vep)\,X_{n_2n_3}^+
(\veq'-\vep)}{E-E_{n_2n_3}(\vep)} \label{20}\end{equation}
where $X_{n_2n_3} (\veQ) =\frac{M_\omega }{\sqrt{N_c}}\,
\psi^{(n_2)}_{Q\bar{q}} (\veQ)\, \psi^{(n_3)}_{\bar{Q}q} (\veQ).$
Now the one-channel Hamiltonian $H_0$ in sector I is augmented by
the term $V_{121}$,
\begin{equation} H=H_0+V_{121},~~~ H\Psi_{n_1}= E\Psi_{n_1}\label{21}\end{equation}
Note, that the coupled-channel (CC) interaction can  be strong
enough to support its own bound states, as was studied in
\cite{Badalian:1981xj}, where this type of resonances was called
the CC resonances.

Let us now turn to the relativistic string Hamiltonian (RSH)
$H_0$, derived from the gauge-invariant meson Green's function in
QCD in  the one-channel case in
\cite{Dubin:1993fk,*Badalian:2008sv}. This Hamiltonian has been
successfully applied to light mesons \cite{Badalian:2002rc},
heavy-light mesons
\cite{Badalian:2007km,Kalashnikova:2001ig,*Badalian:2007yr,*Badalian:2009cx},
and heavy quarkonia
\cite{Badalian:2008bi,*Badalian:2004gw,*Badalian:2008dv,*Badalian:2007zz,Badalian:2008ik,*Badalian:2009bu,*Badalian:2004xv}
and has  a  simple form:

\begin{equation}\label{H_0}
 H_0=\frac{\omega_1}{2} +\frac{\omega_2}{2} +\frac{m^2_1}{2\omega_1}+
 \frac{m^2_2}{2\omega_2} +\frac{\vep^2}{2\,\omega_{\rm red}} +V_{11}(r),
\end{equation}
\begin{equation}\label{V_11}
    V_{11}(r)=\,V_{\rm B}(r)+V_{SD}(r,\omega_i).
\end{equation}
In general, the quantity $\omega_i$ appearing in this expression
is
 an operator, which  in the so-called einbein approximation is   defined by an extremum condition $ \frac{\partial M}{\partial\omega_i}=0$.
  A simple expression for
 the spin-averaged mass $M(nl)$ follows from the RSH (\ref{H_0})
\begin{equation}
 M(nl)=\frac{\omega_{1}}{2}+\frac{\omega_{2}}{2}+\frac{m_1^2}{2\,\omega_1}
 +\frac{m_2^2}{2\,\omega_2} + E_{nl}(\omega_{\rm red}).
\label{M(nl)}
\end{equation}
%
%\end{document}
Here, the excitation energy $E_{nl}(\omega_{\rm red})$ depends on
the reduced mass $\omega_{\rm
red}=\frac{\omega_1\omega_2}{\omega_1+\omega_2}$. The formula
(\ref{M(nl)}) does not contain any additive constant;  for a light
quark (e.g. in  the $D$-meson)   a negative (not small)
nonperturbative self-energy term appears, proportional to
$(\omega_u)^{-1}$;  it has to be added to their masses
\cite{Simonov:2001iv,*DiGiacomo:2004ff}. In the case of charmonium
this term is small; the variables $\omega_i(nl)$, the excitation
energy $E_{nl}(\omega_{\rm red})$, and the w.f. are calculated
from the Hamiltonian (\ref{H_0}) and two extremum conditions
$\partial\,M(nS)/\partial\omega_i=0~(i=1,2)$,
\cite{Dubin:1993fk,*Badalian:2008sv,Simonov:1999qj}:
\begin{eqnarray}\nonumber
&& H_0\,\varphi_{nl}(r)=M(nl)\,\varphi_{nl}(r),\\
&& \omega_i^2(nl)= m_i^2 - \frac{2\,\omega^2_i(nl)\, \partial
E(nl, \omega_{\rm red})}{\partial \omega_i(nl)}\,,\;  (i=1,2).
\end{eqnarray}
The potential $V_B(r)$ in (\ref{V_11})  is derived in the
framework of the Field Correlator Method
\cite{Dosch:1987sk,*Dosch:1988ha,*Simonov:1987rn,DiGiacomo:2000va,Badalian:2008bi,*Badalian:2004gw,*Badalian:2008dv,*Badalian:2007zz}
and is the sum of a pure scalar confining term and a
gluon-exchange part,
 \begin{equation}\label{V_B}
    V_B(r)=\sigma\, r-\frac{4}{3}\frac{\alpha_B(r)}{r}\,,
 \end{equation}
where the vector coupling $\alpha_B(r)$ is taken in the two-loop
approximation and possesses two important features: the asymptotic
freedom behavior at small distances, defined by the QCD constant
$\Lambda_B(n_f)$ [which is considered to be known, because
$\Lambda_B$ is directly expressed via the QCD constant
$\Lambda_{\overline{MS}}(n_f)$ in the $\overline{MS}$
renormalization scheme]; it freezes at large distances. Details
about the effective fine-structure constant can be found in Ref.
\cite{Badalian:2008bi,*Badalian:2004gw,*Badalian:2008dv,*Badalian:2007zz}.

\section{RESONANCES IN  THE DECAY SECTOR}
\label{sect.3}

As it was discussed in \cite{Badalian:1981xj}, the   situation of
two coupled   sectors I, II, $Q\bar Q$ and $(Q\bar q) (\bar Q q)$,
can be treated in two ways:

1) as a coupled system of matrix Green's functions, $$ G^{\,ab},
~~ a,b = I,II$$

2) as a reduction of two-sector problem to the one-sector problem
with energy-dependent ``potential'' $V_{121}$ or
$V_{212}$\footnote{Note, that a parallel treatment of the open
channel problem in nuclear reactions is developed by H.Feshbach
with the help of the projection operators in  his unified theory
of nuclear reactions \cite{Feshbach:1962ut}.}. We shall continue
our one-channel treatment from the point of view of Sector II. In
the same way as it was done before, one can define the
``potential'' $V_{212} \equiv V_{n_2 n_3,\, n'_2n'_3}$
\begin{equation}
V_{n_2n_3,\,n'_2n'_3}(\vep,\,\vep',\, E)=\sum_n
\frac{J_{nn_2n_3}^+(\vep)\, J_{nn'_2n'_3}(\vep')}{E-E_n}\,.
\label{23}\end{equation}
Defining also $ J_{nn_2n_3} (\ver) \equiv
\int\frac{d^{\,3}\vep}{(2\pi)^3}\, e^{i\,\vep \ver}\, J_{nn_2n_3}
(\vep),$ one can write
\begin{equation}
V_{n_2n_3,\,n'_2n'_3}(\ver,\,\ver',\, E)=
\sum_n\frac{J_{nn_2n_3}^+(\ver)\,
J_{nn'_2n'_3}(\ver')}{E-E_n}\label{24}\end{equation}
and as a result one obtains a system of equations in sector II
\begin{equation} \big(H_0 + V_{22} (\ver)\big)\, \psi_{n_2n_3} (\ver) +\int
V_{n_2n_3,\,n'_2n'_3} (\ver,\,\ver',\,E)\,
\psi_{n'_2n'_3}(\ver')\, d^{\,3}\ver'=
E\,\psi_{n_2n_3}(\ver)\label{25}\end{equation}
where $V_{22} (\ver)$ is a direct interaction between two color
singlet mesons, which we neglect in the first approximation, $H_0$
is the same as in (\ref{H_0}) but for $m_1, m_2$ equal to masses
of mesons with quantum numbers $n_2, n_3, H_0=H_0(n_2n_3)$.
Neglecting $V_{22}$, one can easily rewrite (\ref{25}) for the
separable interaction (\ref{24})
\begin{equation}
\psi_{n_2n_3} (\ver) =- \sum_n \int
d^{\,3}\ver'\,d^{\,3}\ver^{\pp}\, G_{n_2n_3}^{(0)}(\ver,\ver')\,
\frac{J_{nn_2n_3}^+(\ver')\,
J_{nn'_2n'_3}(\ver'')}{E-E_n}\,\psi_{n'_2n'_3}(\ver^{\pp})\,,
\label{26}\end{equation}
where
\begin{equation}
G^{(0)}_{n_2n_3} (\ver, \ver') =\int \frac{d^{\,3}\vek}{(2\pi)^3}
\frac{e^{i\,\vek(\ver-\ver')}}{H_0^{(n_2n_3)}(\vek)-E}\,.
\label{27}
\end{equation}
Introducing $\varphi_n\equiv \int J_{nn_2n_3}(\ver)
\psi_{n_2n_3}(\ver)\,d^{\,3}\ver$, and integrating both sides of
(\ref{26}) with $J_{mn_2n_3}(\ver)\,d\ver$, one has  from
(\ref{26})
\begin{equation}
\varphi_m =\sum_n \frac{w_{mn}(E)\, \varphi_n}{E-E_n} \label{28}
\end{equation}
with the same $w_{mn}(E)$ as in (\ref{9}), and the equation for
eigenvalues is again (\ref{12}).

Since the  CC  interaction (\ref{24}) is separable, one can study
the structure of the spectrum of our CC problem in more detail; in
particular, whether there can appear poles (CC resonances in
terminology of \cite{Badalian:1981xj}) due to strong CC
interaction, which are additional to the one-sector spectrum of
poles $E_n$, the latter being  simply shifted by CC. As was argued
in \cite{Badalian:1981xj}, we define the integral
\begin{equation}
I_{n_2n_3} (E) =\left| \int
\frac{d^{\,3}\vep}{(2\pi)^3}\sum_n\frac{
|J_{nn_2n_3}(\vep)|^{\,2}}{\big(E-E_n\big)
\big(H_0^{(n_2n_3)}(\vep) -E\big)}\right|. \label{29}
\end{equation}
According to \cite{Badalian:1981xj},  a bound state in a single
channel $n_2n_3$ due to CC with  the sector I  can exist,  if in
the region, where (\ref{29}) is real (below threshold $E_{th}(n_2,
n_3)$), it becomes larger than one
\begin{equation}
I_{n_2n_3} (E) >1,~~ E< E_{th} (n_2,n_3). \label{30}
\end{equation}
In the momentum space one has
\begin{equation}
\tilde H_0(\vep)\,\psi_{n_2n_3}(\vep) + \int V_{n_2n_3,\,
n'_2n'_3}(\vep,\,\vep',\, E)\,\psi_{n'_2n'_3}(\vep')\,
\frac{d^{\,3}\vep'}{(2\pi)^3} = E\,\psi_{n_2n_3}(\vep), \label{31}
\end{equation}
which yields the same equation  as in (\ref{28}). As before in
Eq.\,(\ref{12}),  one obtains from (\ref{31})  the equation $\det
\big(E-\hat E_n-\hat w\big) =0$, which defines all poles in the
cut $E$ plane below thresholds and on the second, and higher
Riemann sheets.

\section{THEORY OF COUPLED-CHANNEL RESONANCES BASED ON THE WEINBERG EIGENVALUE METHOD}
\label{sect.4}

The coupled-channel problem can be quantified using the eigenvalue
analysis introduced by Weinberg
\cite{Weinberg:1963zz,*newton:319,*Smithies1958}. Although this
formalism has been developed long ago, it is still not widely
known. That is why in this section we present a short summary of
corresponding formulae leaving details to the Appendix D.

The Schrodinger equation for two-body like Hamiltonian $H = H_0 +
V$ can be written in the standard (time-independent) way
\begin{equation}\label{35d}
\big(H_0-E\big)\,\Psi_E(r)=-V\,\Psi_E(r),
\end{equation}
where E is a spectral variable,  $\Psi_E(r)$ is an energy
eigenstate and $V$ is an operator, which in the nonlocal case acts
in (\ref{35d}) as $V(\Psi_E(\ver))=\int V(\ver,
\ver')\Psi_E(\ver')\, d\ver'$. In the Weinberg method, instead,
w.f. are the eigensolutions of
\begin{equation}\label{}
    \big(H_0-E\big)\,\Psi_\nu(r,E)=\frac{-V}{\eta_{\nu}(E)}\,\Psi_\nu(r,E)
\end{equation}
where E is the continuous  parameter  entering w.f. and the index
$\nu$ labels the discrete eigenvalues and eigenvectors. The
Weinberg eigenvalue $\eta_\nu(E)$ is the potential scale and thus
the spectrum consists of all the potential rescalings that give
solution to that equation, for given energy $E$.

Let us now turn to the question of the rigorous definition of the
resonance wave function and start with the one-state situation,
when only  one state is considered in sector II, with fixed $n_2,
n_3$. The induced interaction $V_{212}(\ver, \ver', E)$ has the
form (\ref{24}) and direct interaction $V_{22}$ is neglected for
simplicity. One can exploit the WEM
\cite{Weinberg:1963zz,*newton:319,*Smithies1958}, and divide the
potential $V_{212}(\ver, \ver', E)$ in (\ref{29}) by an
energy-dependent factor $\eta_\nu(E)$ considering instead of Eq.
(\ref{25}) another one (with $V_{22}\equiv 0)$
\begin{equation} H_0\,\Psi_\nu(\ver, E) +
\int\frac{V_{212}(\ver,\,\ver',\,E)}{\eta_\nu(E)}\,
\Psi_\nu(\ver', E)\, d^{\,3}\ver' = E\,\Psi_\nu(\ver,
E)\,,\label{32}\end{equation}
which defines for each $E$ the eigenvalue $\eta_\nu(E)$ and
eigenfunction $\Psi_\nu(r, E)$, with the boundary conditions
\begin{equation} \Psi_\nu (0) ={\rm const},~~ \Psi_\nu (r\to \infty) =
C\,\frac{e^{i\, kr}}{r},~~ k=\sqrt{2\tilde
M(E-E_{th})}.\label{33}\end{equation}
For $E<E_{th}$ one has instead $\Psi_\nu(r\to \infty) \sim
c\,\exp(-\kappa r)/r.$ In  the WEM, resonance structures as well
as bound sates can be obtained in terms of Weinberg eigenvalues
$\eta_\nu (E)$. Note, that a solution of integro-differential
equation (\ref{32}) in the coordinate space can satisfy these
boundary conditions for each energy $E$ only  for some discrete
value of $\eta_\nu(E),\, \nu=1,2,...$ . Compare e.g. with  the
case of bound states (energy below threshold), where boundary
conditions at origin and infinity can be matched only for discrete
energy $E_i$ in standard formalism ($\eta=1)$ or at some
$\eta_i(E)$ for any $E$ in WEM, with the relation $\eta_i(E_i)=1$.

The normalization of wave functions is
\begin{equation}
\int d\ver\,d\ver'\,\Psi_\nu(\ver, E)\, \hat V_{212}(\ver, \ver',
E)\,\Psi_{\nu'}(\ver, E) = - \delta_{\nu\nu'}\, \eta_\nu (E).
\label{34}
\end{equation}
Note, that $\hat V_{212}(E)$ is real analytic (holomorphic) for all $E$ except
for  pole positions, and  the off-shell $t$-matrix looks as (see
\cite{Weinberg:1963zz,*newton:319,*Smithies1958} and Appendix D for a
derivation)
\begin{equation}
t(\vep,\,\vep',\,E)= - \sum_\nu
\frac{\eta_\nu(E)}{1-\eta_\nu(E)}\, a_\nu(\vep,\,E)\,
a_\nu(\vep',\,E), \label{36}
\end{equation}
with
\begin{equation}
a_\nu(\vep,\,E) =\big(H_0(\vep) -E\big)\, \Psi_\nu(\vep,\,E),
\label{37}
\end{equation}
\begin{equation}
\int \frac{a_\nu (\vep,\,E)\, a_{\nu'} (\vep,E)}{H_0 (\vep) -E}
\frac{d^{\,3} \vep}{(2\pi)^3}=\delta_{\nu\nu'}, \label{38}
\end{equation}
and the Green's function (\ref{11}) has the form

\begin{equation}\label{38*}
\nonumber G_{Q\bar{Q}}^{(I)}(1,2;E)=\sum_\nu \frac{\Psi_\nu(1,E)\,
\Psi^+_\nu(2,E)}{1-\eta_\nu(E)}\,.
\end{equation}

The sum over $\nu$ is fast  converging as one can see from the
example of square well and other potentials fast decreasing at
$\infty$
\cite{Narodetsky:1980nx,*Narodetsky:1969jf,*Herzenberg,*Fuda}.
Therefore in what follows in our calculations we shall consider
only one term in the sum over $\nu$, which is relevant for a given
threshold.

The  Breit-Wigner resonances in sector II are obtained from the
condition that for some $\nu=\nu_0$, $\eta_{\nu_0}\left(E_0 -
\frac{i\Gamma}{2}\right) =1$, and
\begin{equation}
\eta_{\nu_0}(E) =1 + \eta'_{\nu_0}\big(E_0 -
\frac{i\Gamma}{2}\big)\left(E-E_0 + \frac{i\Gamma}{2}\right)+...
\label{39}
\end{equation} Note, that
the corresponding $\Psi_{\nu_0} (r, E)$ serves as the normalized
resonance wave function and can be used e.g. to calculate average
values of some operator or perturbative shift of resonance
position. The $Q\bar{Q}$ Green's function with account of channel
coupling can be written near the pole $E=E^R$ (resonance) as

\begin{equation}\label{39star}
G_{Q\bar{Q}}^{(I)}(1,2;E)=\frac{\bar{\Psi}_\nu(1,E)\,\bar{\Psi}^+_\nu(2,E)}{E^R-E},\quad
\bar{\Psi}_\nu=\frac{\Psi_\nu}{\sqrt{\frac{d \eta_\nu(E^R)}{dE}}}
\end{equation}

As it is seen from (\ref{32}), the introduction of WEM eigenvalue
in equations reduces to the replacement $w_{mn} (E)\to
\frac{w_{mn}(E)}{\eta (E)}$, hence the resulting equation for the
calculation of $\eta (E)$ is
\begin{equation}
\det \left( \hat 1-\frac{\hat w(E)}{\eta(E)} \frac{1}{E-\hat
E}\right) =0 ,~~ (\hat E)_{mn} = E_n\, \delta_{mn}. \label{40}
\end{equation}
Eq.(\ref{40}) is of the $n$-th power in $\eta$, when $n$ levels $
(Q\bar Q)$ are taken into account, and this yields $n$ roots
$\eta_k(E)$, $k=1,...n$. Total number of poles is given by
solutions $\eta_k (E_l) =1$, $l=1,...N_p$, where $N_p$ depends on
behavior of $\eta_k(E)$.

We started formally with the one channel in sector II, i.e. with
fixed $n_2,n_3$, hence in $\hat w(E)$ in Eq.(\ref{40}) the sum
over $n_2,n_3$ (cf. Eq.(\ref{9})) reduces to one term. However,
for several states $n_2,n_3$ one has equation (\ref{25}) with
interaction kernel $\hat V_{212}$ as a matrix in indices
$n_2,n_3,\,n'_2,n'_3$, and if $\eta (E)$ in (\ref{32}) does not
depend on $n_2,n_3$, then as a result one has for $\eta(E)$ the
same Eq.(\ref{40}), but now with $\hat w (E)$, which corresponds
fully to (\ref{9}), i.e. contains the sum over $n_2,n_3$.

%The $S$-matrix (\ref{35}) in this case refers to the  ``compound
%state'' of hadron-hadron system, with partial widths for each
%specific channel $n_2,n_3$.

Let us discuss how the basic equation (\ref{40}) changes for many
channels $n_2,n_3$. We start with Eq. (\ref{26}), which is
equivalent to (\ref{32}), when one introduces in (\ref{26}) in the
denominator on the r.h.s. the factor $\eta_\nu(E)$.  Multiplying
both sides of this modified Eq. (\ref{26}) with $J_{m n_2
n_3}(\ver)$ and integrating and summing over $n_2,n_3$ one obtains
equation similar to (\ref{28})
\begin{equation} \varphi^\nu_m (E)=\frac{1}{\eta_\nu(E)}\sum_n
\frac{w_{mn}(E)\, \varphi^\nu_n (E)}{E-E_n}\,,
\label{41}\end{equation}
where $\varphi_m (E) = \sum\limits_{n_2n_3} J_{mn_2n_3} (\ver)\,
\psi_{n_2n_3}(\ver)\, d^{\,3} \ver$, and $w_{mn}(E)$ is the same,
as in (\ref{9}), i.e. again with the sum over $n_2, n_3$. The
resulting equation to determine $\eta(E)$ is again (\ref{40}), and
all equations (\ref{32})-(\ref{34}) have  the same form, if one
takes into account, that $\Psi_\nu$ is a column of $\psi_{n_2n_3}$
components and $\hat V_{212} $ is a matrix in indices  $n_2n_3,\,
n'_2n'_3$.

\begin{figure}[t]
\begin{minipage}[h]{0.45\linewidth}
\center{\includegraphics[angle=0,width=0.8\textwidth]{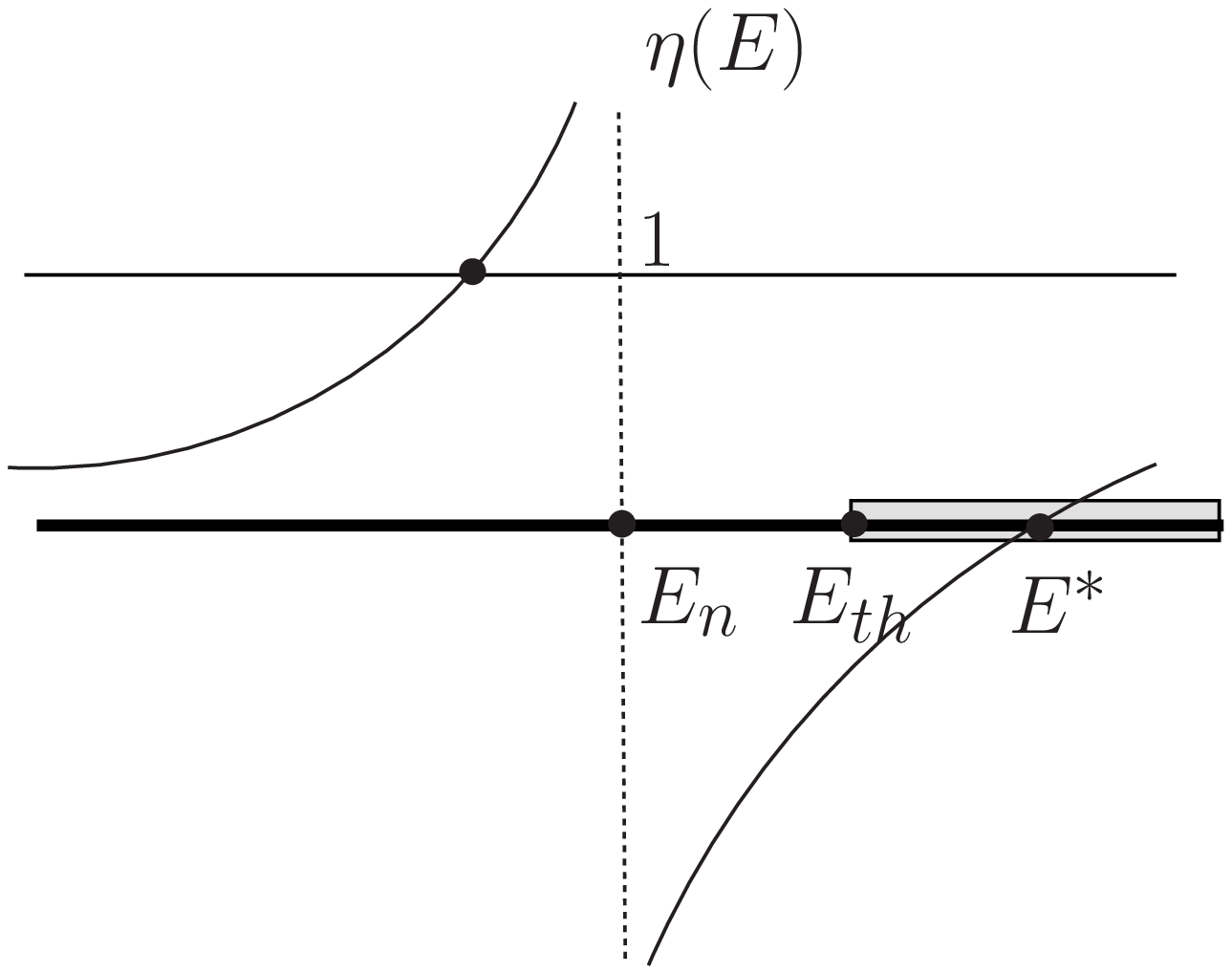} \\
(a) case $E_n<E_{th}$}
\end{minipage}
\hfill
\begin{minipage}[h]{0.45\linewidth}
\center{\includegraphics[angle=0,width=0.8\textwidth]{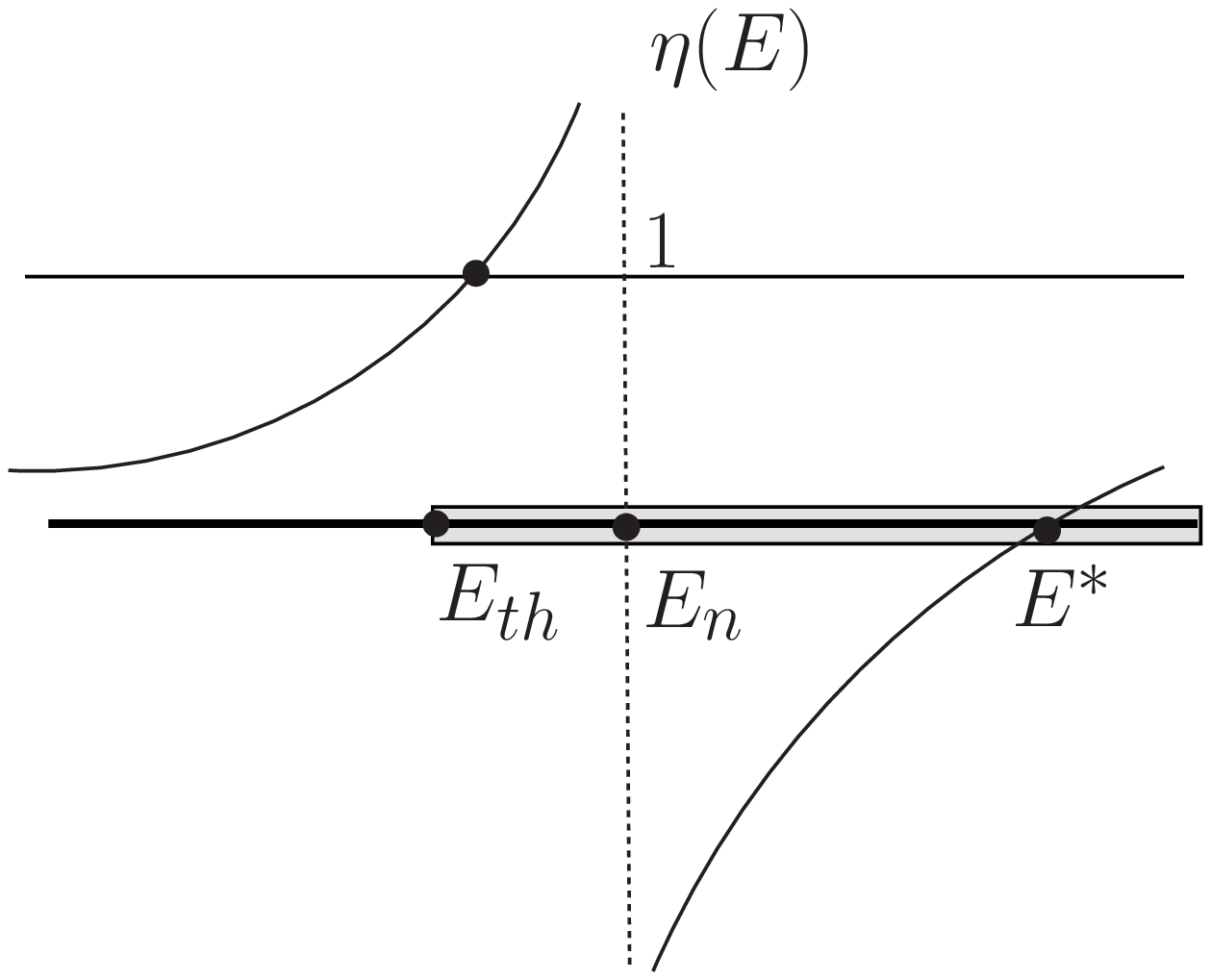} \\
(b) case $E_{th}<E_n<E^*$ }
\end{minipage}
\begin{minipage}[h]{0.45\linewidth}
\center{\includegraphics[angle=0,width=0.8\textwidth]{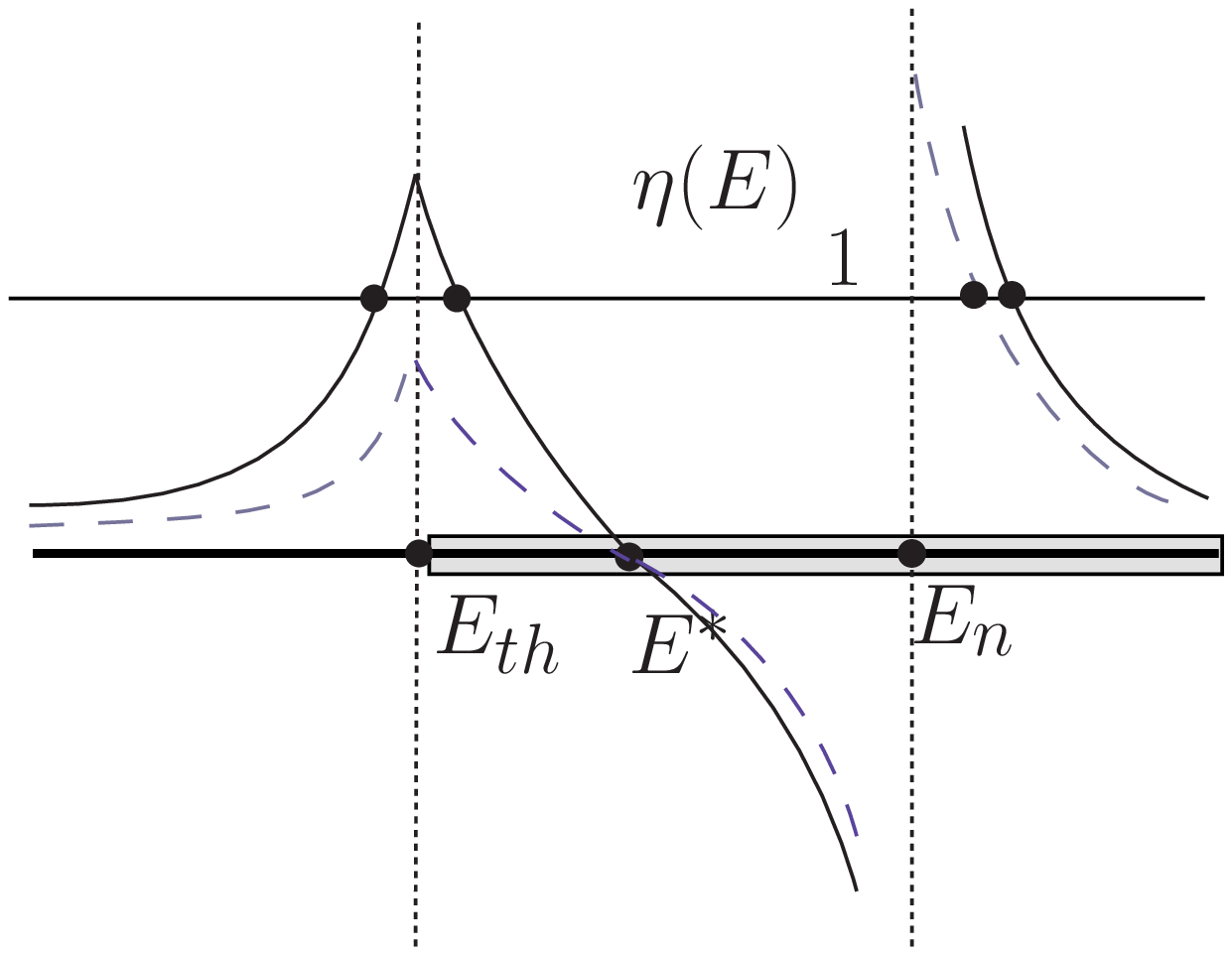} \\
(c) case $E_n>E^*$}
\end{minipage}
\hfill
\begin{minipage}[h]{0.45\linewidth}
\center{\includegraphics[angle=0,width=0.8\textwidth]{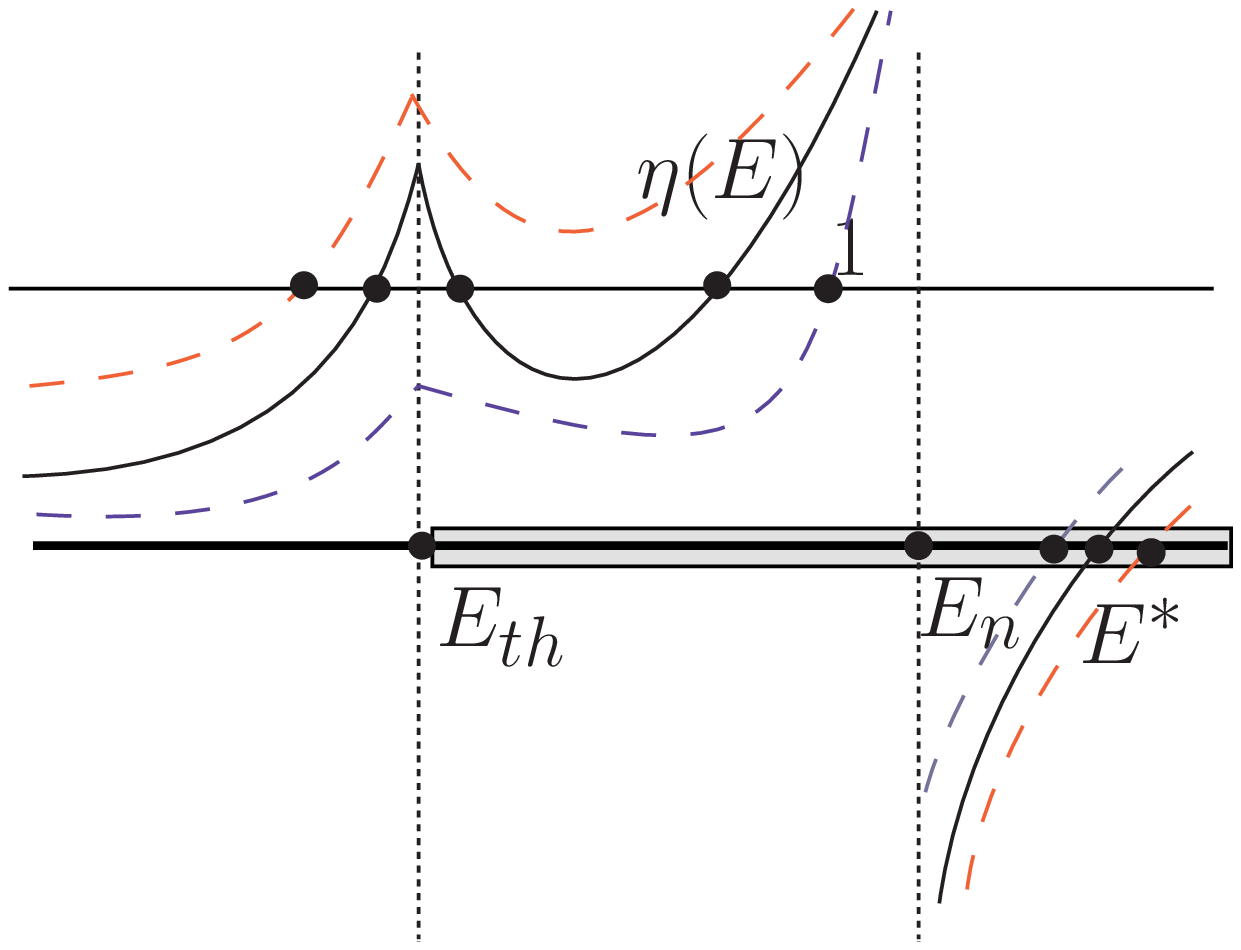} \\
(d)  case $E_{th}<E_n<E^*$ }
\end{minipage}
 \caption{Qualitative pictures of Weinberg eigenvalues $\textrm{Re}\,(\eta(E))$ as a function of energy $E$, where
 $E_n$ is the eigenvalue of the single-channel relativistic string Hamiltonian $H_0$ (bare resonance), $E_{th}$
 denotes threshold and $E^*$ is the point, where $\textrm{Re}(w_{nn}(E))$ changes sign.} \label{Fig.2}
\end{figure}

Finally, the separate components $\psi_{n_2n_3} $ are found
through $\varphi_n (E)$ via (cf. (\ref{26}))
\begin{equation} \psi_{n_2n_3} (\ver) = - \sum_n
\frac{\varphi^\nu_n (E)}{\eta_\nu(E)\, (E-E_n)} \int
G^{(0)}_{n_2n_3} (\ver,\,\ver')\,
J^+_{nn_2n_3}(\ver')\,d\ver',\label{42}
\end{equation}
and partial  widths of the resonance are
found in lowest approximation as (for one channel $n$ in sector I)
\begin{equation}
\Gamma_{nn_2n_3} (E^R)= 2\,\textrm{Im}_{n_2n_3} \big(w_{nn}
(E^R)\big)=2\pi\int \frac{d^{\,3}\vep}{(2\pi)^3}
\,|J_{nn_2n_3}(\vep)|^{\,2}\,\delta\big(E^R -E_{n_2n_3}
(\vep)\big).\label{43}
\end{equation}
To understand the possible origin and position of resonances in
our CC problems, one can consider  several typical cases,
depending on relative positions of bare resonances $E_n$ and
thresholds $E_{th}(n_2n_3)$. Consider first  one state in sector
I, one  state in sector II, then $w_{nn} (E) <0$ for $E<E_{th}$,
and $\textrm{Re}\, \big(w_{nn}(E)\big)$ changes sign at $E=E^*$.
The   resulting  qualitative picture of $\eta(E)= \frac{w_{nn}
(E)}{E-E_n}$ is shown in Figure \ref{Fig.2} for three cases: $E_n<
E_{th}$ (Fig. \ref{Fig.2}(a)); $E_{th}< E_n < E^*$ (Fig.
\ref{Fig.2}(b,d)); $E_n>E^*$ (Fig.\ref{Fig.2}(c)). In
Fig.\ref{Fig.2}(a,b) one can see one critical energy for which
$\eta(E)=1$. This point corresponds to the shifted energy level
$E_n$.

Note  the possibility of a pair of additional roots of equation $
\eta(E) =1$, when $E_n> E^*$, $E_{th}< E_n < E^*$ and
\begin{equation} w_{nn} (E_{th})
> |E_n- E_{th}|.\label{44}\end{equation}
The condition (\ref{44}) defines the strength of CC interaction in
the situation depicted in Fig.\ref{Fig.2}(c,d) which is necessary
 one additional pole near the threshold energy (see Appendix E
for details). As we shall show below in section \ref{sect.8}, the
situation of Fig.\ref{Fig.2}(d) is most likely realized in the
$~^3P_1$ state of charmonium, where the threshold peak corresponds
to the $X(3872)$ state. In this case actually two close-by
thresholds are present ($D_0D_0^*$ and $D_+D^*_-$), and as will be
seen, the experimentally observed situation with one peak at
lowest threshold and wide structure near $E\sim3.940$ GeV indeed
occurs.

Consider now the case of two  levels  in sector I, $E_1$ and
$E_2$; and one (or more)  state in sector II. The equation for
$\eta(E)$ has the form
\begin{equation}
\eta^2(E) -\eta(E)\, (\tilde w_{11} + \tilde w_{22}) - \tilde
w_{12}\, \tilde w_{21} =0, \label{45}
\end{equation}
with $\tilde w_{ik}\equiv\frac{w_{ik}(E)}{E-E_k}$, and the result
\begin{equation}
\eta_\pm (E) =\frac12\, (\tilde w_{11}+\tilde w_{22})\pm \frac12\,
\sqrt{ (\tilde w_{11} -\tilde w_{22})^2 + 4\tilde w_{12}\, \tilde
w_{21}}. \label{46}
\end{equation}
where for notational convenience we have suppressed the energy
dependence of the $\tilde w_{nm}$.

Near $E=E_1,~ \eta_\pm (E)$ can be identified with the one-channel
eigenvalues $\eta_1(E) \equiv
 \frac{w_{11}(E)}{E-E_1}$ and $\eta_2 (E)
 =\frac{w_{22}(E)}{E-E_2}$, namely for $E<E_1$ and $E\to E_1$ one
has
\begin{equation}
\eta_+ (E\to E_1) = \eta_1(E) +
\frac{w_{12}(E)\,w_{21}(E)}{w_{11}(E)\,(E-E_2)}+... \label{47}
\end{equation}

\begin{equation}
\eta_- (E\to E_1) = \eta_2(E) -
\frac{w_{12}(E)\,w_{21}(E)}{w_{11}(E)\,(E-E_2)} \label{48}
\end{equation}
and for $E\to E_2$ one should change in (\ref{47}), (\ref{48})
$1\leftrightarrow 2$.

The situation with trajectories $\eta_\pm (E)$ is in general
rather complicated, and we describe below only one case when $E_1<
E_2< E^*_{ik}, i,k=1,2$ where
$\textrm{Re}\,\big(w_{ik}(E^*_{ik})\big)=0$, and in case of strong
mixing of channels 1 and 2 the point $E_0$, where
$\textrm{Re}\,\big(\eta_+ (E_0)\big) = \textrm{Re}\,\big(\eta_-
(E_0)\big)$ lies between $E_1$ and $E_2$. (The position of
$E_{th}$ is irrelevant for  the situation where all imaginary
parts are neglected). However for weak mixing of channels 1,\,2
roots $\eta_+, \eta_-$ never coincide. One can see from (\ref{46})
that in the weak mixing case only two  poles remain, corresponding
to shifted levels $E_1,\,E_2$ and no new resonances appear at
least for $E<E^*_{ik}$.

%For strong coupling, i.e. when $w_{12}$ and $w_{21}$ are large
%enough, so that two roots $\eta_\pm (E)$ can coincide, the
%situation is depicted in Fig.3 (b), solid lines; the broken curve
%for $\eta$-  is for the case of strong mixing, when $w_{22}(E)
%-\frac{w_{11} w_{21}}{w_{11}}$ is positive in that region. It is
%conceivable,  that the most interesting case is (as in the
%one-channel situation) when $E_1> E^*_{ik}$.

\section{MIXING OF STATES IN THE WEINBERG FORMALISM}
\label{sect.5}

The WEM  solves the important problem of constructing the full set
of orthogonal states in the coupled channel problem, and thus the
problem of mixing  of states. This is nontrivial in  the situation
under investigation, since the interaction in the sector I induced
by the coupling to the sector II, $V_{121}(\ver, \ver')$, Eq.
(\ref{17}), is energy dependent and hence violates the
orthogonality of eigenstates. In addition, for energies above
threshold, this interaction is complex and makes the corresponding
states the resonances, which cannot be normalized and
orthogonalized to each other in the ordinary way. Happily, WEM
allows to define all states and their  mixing in the
mathematically rigorous way, as we shall now show.

We  start with the formulation in sector I and  write starting
from (\ref{21}) the WEM equation
\begin{equation}
H_0\,\Psi_\nu(\ver, E) +
\int\frac{V_{121}(\ver,\,\ver',\,E)}{\eta_\nu(E)}\,
\Psi_\nu(\ver', E)\, d^{\,3}\ver' = E\,\Psi_\nu(\ver, E)\,,
\label{49}
\end{equation}
while  the unperturbed states  $\Psi_n(\ver) $ satisfy
\begin{equation}
H_0 \Psi_n(\ver) = E_n \Psi_n(\ver). \label{50}
\end{equation}
Note, that $\Psi_\nu(\ver,E)$ depend on energy $E$, while
$\Psi_n(\ver)$ do not. Similarly to (\ref{34}), the orthogonality
condition is

\begin{equation} \int d\ver\, d\ver' \Psi_\nu(\ver, E)\, V_{121} (\ver, \ver',
E)\, \Psi_{\nu'} (\ver', E) =-\, \delta_{\nu\nu'}\, \eta_\nu
(E).\label{51}\end{equation}

Consider now the  expansion of  a WEM state in the  set of
$\Psi_n$ states,
\begin{equation}
\Psi_\nu (\ver, E) = \sum_n c_n^{\nu} (E)\, \Psi_n
(\ver).
\label{52}
\end{equation}
Taking into account, that
\begin{equation}
\int d\ver\, d\ver'\, \Psi_n (\ver)\, V_{121}(\ver, \ver', E)\,
\Psi_m (\ver') = w_{nm} (E), \label{53}
\end{equation}
and multiplaying both sides of (\ref{49}) with $\Psi_{\nu'} (\ver,
E)$ and integrating over $d\ver$, one obtains
\begin{equation}
\sum_n c_n^{\nu'}(E)\, c_n^\nu(E)\, \big(E_n-E\big) +\sum_{m,n}
\frac{c_n^{\nu'}(E)\, w_{nm}(E)\, c_m^\nu(E)}{\eta_\nu(E)} =0\,.
\label{54}
\end{equation}
Thus one obtains the equation for eigenvalues $\eta_\nu(E)$
\begin{equation}
\det \left(\hat E-E+ \frac{\hat w}{\eta_\nu(E)}\right)=0
\label{55}
\end{equation} which coincides with
(\ref{40}), obtained in sector II. Now we are specifically
interested in the coefficients $\{ c_n^\nu\}$, $\{ c_n^{\nu'}\}$
for two different eigenvalues $\eta_\nu (E),\, \eta_{\nu'}(E)$.

The first condition follows from (\ref{51}), (\ref{54})
\begin{equation}
\sum_n c_n^{\nu'}(E)\,c_n^\nu(E)\,\big(E_n-E\big)
=\delta_{\nu\nu'}. \label{56}
\end{equation}
It is convenient to introduce reduced coefficients:
\begin{equation} c_n^\nu(E) = \frac{\bar c^\nu_n(E)}{\sqrt{E_n-E}};~~ \bar w_{mn}
(E)
=\frac{w_{mn}(E)}{\sqrt{(E_m-E)(E_n-E)}}.\label{57}\end{equation}
Then the solution for two eigenvalues in (\ref{55}) is
\begin{equation}
\eta_\nu (E) = \frac{-\big((\bar w_{22} +\bar w_{11}) \pm
\sqrt{(\bar w_{22} +\bar w_{11})^2 -4 \det \bar w}\,\big)}{2}
\label{58}
\end{equation}
here and further for notational convenience we will suppress the
energy dependence of the $\bar w_{nm}$. The normalization
condition has the form
\begin{equation}
\sum_n \bar c_n^\nu(E)\, \bar c_n^{\nu'}(E) = \delta_{\nu\nu'},
~~\sum_{n,m} \bar c^\nu_n(E)\, \bar w_{nm}(E)\,
c^{\nu'}_m(E)=-\delta_{\nu\nu'}\, \eta_\nu(E) \label{59}
\end{equation}
Let us take one concrete example of two states in the subthreshold
region (e.g. ($2^{\,3}S_1)$ and ($1^3D_1)$ states of charmonium,
however at this stage they  are not specified).

Keeping only two states $n=1,2$ e.g. for $(2^{\,3}S_1)$ and
$(1^{\,3} D_1)$, one can write for $c_n^\nu(E)$,
$\nu=\alpha,\,\beta $
\begin{equation}
\bar c_1^\alpha(E) =\cos \varphi(E),~ \bar c_2^\alpha(E) =\sin
\varphi(E);\quad \bar c_1^\beta(E)=\sin \varphi(E),~ \bar
c_2^\beta(E)=-\cos\varphi(E). \label{60}
\end{equation}
Note, that the appearance of $O(2)$ coefficients is not accidental
since $w_{nm}(E)$ is symmetric in $n,m$.

We are thus e.g. looking for the shifted and mixed $(2^{\,3}S_1)$
state, denoted by $\alpha$, and the same for $(1^3 D_1)$ state,
denoted by $\beta$.

$$\Psi^\alpha(E) = \frac{\cos\varphi(E)}{\sqrt{E_1-E}}\,\,\Psi_1 +
\frac{\sin\varphi(E)}{\sqrt{E_2-E}}\,\,\Psi_2$$
\begin{equation}
\Psi^\beta(E) = \frac{\sin\varphi(E)}{\sqrt{E_1-E}}\,\,\Psi_1-
\frac{\cos\varphi(E)}{\sqrt{E_2-E}}\,\,\Psi_2\,. \label{61}
\end{equation}
To find $\cos \varphi(E)$, one can use the second equation in
(\ref{59}), which yields
$$\sin^2\varphi(E)\, \bar w_{11} -\cos\varphi(E)\,\big(\bar w_{12}+\bar
w_{21}\big)+\cos^2\varphi(E)\, \bar w_{22} =-\eta_\beta(E)$$
\begin{equation}
\cos^2\varphi(E)\, \bar w_{11} -\cos\varphi(E)\, \big(\bar
w_{12}+\bar w_{21}\big)+\sin^2\varphi(E)\, \bar w_{22}
=-\eta_\alpha(E). \label{62}
\end{equation}
This gives the condition $\bar w_{11}+\bar w_{22}=-
\,\big(\eta_\alpha(E) +\eta_\beta(E)\big)$, which is identically
satisfied, and the final result for $\cos^2\varphi(E)$
\begin{equation} \cos^2 \varphi(E) =\frac{\bar w_{11} -\bar w_{22}+D}{2D},~~
D=\sqrt{(\bar w_{11} -\bar w_{22})^2+ 4 \bar w_{12}\bar
w_{21}}.\label{63}\end{equation}

Note, that the sign of $D$ is  connected with the corresponding
choice of the root in (\ref{58}), for $\eta_\alpha(E)$ (lower in
energy state) we have chosen the sign $+$.

It is clear, that $\cos \varphi$ depends on $E$ and therefore to
define finally the mixing coefficient,  one should fix the energy.
E.g. for the state $\beta$, the eigenvalue $\eta_\beta(E)$ crosses
the line $\eta(E)=1$ at the resonance position $E= E^R_\beta$,
complex in general, and the mixing coefficient of interest from
(\ref{60}) is $c_1^\beta =\frac{\sin \varphi
(E^R_\beta)}{\sqrt{E_1-E^R_\beta}}$, while the mixing coefficient
of the state $\alpha$ is to be taken at $E=E^R_\alpha$,
$c^\alpha_2=
\frac{\sin\varphi(E^R_\alpha)}{\sqrt{E_2-E^R_\alpha}}$.

Hence, for small shifts $E^R_\beta \cong E_2, E_\alpha^*\approx
E_1$, and energy independent $\varphi$, one recovers the symmetry
condition \begin{equation} |c_1^\beta| \approx
|c_2^\alpha|.\label{64}\end{equation}

Finally, one should connect normalizations of $\Psi_n$ and
$\Psi^{\alpha,\beta}$. This can be done, if one  considers the
limiting case of one channel $\nu$, where according to (\ref{53}),
(\ref{51}), one has
\begin{equation}
\big(c_n^\nu(E)\,w_{nm}(E)\,c_m^\nu(E)\big) = -\eta_\nu(E)
\label{65}
\end{equation}
and for $E=E^R_\nu$ (at the resonance position),
$\eta_\nu(E^R_\nu)=1$, and for one level $n$ from (\ref{55}) one
has $w_{nm} (E^R_\nu)= E^R_\nu - E_n \equiv -\Delta E_n$. Hence in
the one-channel -- one-level limit we have
\begin{equation}
(c^\nu_n)^2 \Delta E_n =1,~~ c_n^\nu =\frac{1}{\sqrt{\Delta
E_n}}.\label{66}
\end{equation}
Therefore  if only one level n is kept, then the normalized WEM
states can be defined as
\begin{equation} \bar \Psi^\alpha(E^R_\alpha) = \Psi^\alpha(E_\alpha^*)
\sqrt{\Delta
E_n},~~\int\big(\Psi^\alpha(E^R_\alpha)\big)^2\,d^{\,3}r=1
\label{67}
\end{equation}
and finally the standard normalized mixing coefficients are
\begin{equation}
\tilde c_1^\beta =\frac{\sin\varphi (E^R_\beta)\, \sqrt{\Delta
E_\beta}}{\sqrt{E_1-E_\beta^*}},~~\tilde c_2^\alpha
=\frac{\sin\varphi (E^R_\alpha)\, \sqrt{\Delta
E_\alpha}}{\sqrt{E_2-E_\alpha^*}}\,, \label{68}
\end{equation}
where $\Delta E_\beta= E_2 -E^R_\beta, ~~ \Delta E_\alpha =E_1 -
E^R_\alpha$. One can see, that in general coefficients are less
than unity due to ratios of square roots. We finally write for
$\sin \varphi$
\begin{equation} \sin^2 \varphi (E) = \left\{ \frac{(\bar w_{11}-\bar w_{22})^2
+ 2\,(\bar w_{12})^2 + 2\,(\bar w_{21})^2- (\bar w_{11} -\bar
w_{22})\,D}{ 2\,D^2}\right\}\label{69}\end{equation}
Another (and physically more motivated) normalization for
$\Psi^\alpha(E_\alpha^*)$ follows from (\ref{39star}), which can
be written as
$$G_{Q\bar{Q}}^{(I)}(1,2;E)=\frac{\Psi_\alpha(1,E^R)\,\Psi^+_\alpha(2,E^R)}{(E^R-E)\,\frac{d\eta_\alpha(E^R)}{dE}}$$
Estimating
$\frac{dw_{nn}(E^R)}{dE}=\frac{w_{nn}(E^R)\,\xi}{|E_\alpha^*-E_{th}|}$,
$\xi<1$, one obtains
$\frac{d\eta_\alpha(E_\alpha^*)}{dE}=\frac{1}{E_n-E_\alpha^*}+\frac{\xi}{|E^R-E_{th}|}$,
and the defacto wave functions is
$\Psi_\alpha(1,E^R)/\sqrt{\frac{d\eta_\alpha(E^R)}{dE}}$, which is
close to (\ref{67}) for $\xi \ll 1$.

\section{RESULTS AND DISCUSSION}
\label{sect.8}

The formalism given in this paper is based on the explicit
knowledge of wave functions in both sectors I and II and yields
the CC interaction operator $\hat w(E)$ expressed via the overlap
integrals, see Eq.(\ref{9}). The resulting effective interaction
in each sector is energy dependent  due to $\hat w(E)$, and
violates usual orthonormality properties for wave functions.
Moreover, new states appear for energies above thresholds, and one
needs a rigorous formalism to treat the complete set of
eigenfunctions for such operators.  The WEM is indispensable for
this purpose. In Eq.(\ref{40}) explicit conditions are written
down for Weinberg eigenvalues $\eta(E)$. It is important, that
$\eta(E)$ has simple analytic properties in the $E$-plane.
Therefore physical quantities expressed via $\eta(E)$, like
scattering amplitude (\ref{70}) or production cross-section
(\ref{71}) have  a definite analytic expression near the pole(s),
different from the Breit-Wigner form in general. This property is
more important in case of the complicated arrangement of
thresholds and poles, as it is in the case of $X(3872)$, see
below.

Another practical advantage of WEM is the complete set of states
for each energy $E$, allowing to define unambiguously symmetric
mixing coefficients, as it was explained in section \ref{sect.5}.

Before a detailed discussion of results, one should stress two
main features of the closed channel pole $E_n$ behavior under the
influence of CC: 1) CC is attractive for all states below CC
threshold; 2) CC is attractive in some region $E_{th} \leq E_n
\leq E^*$ above threshold and repulsive for $E>E^*$ (in the limit
of small width). Both statements follow from the condition
$\textrm{Re}\,\big(w(E)\big)<0$ or $\textrm{Re}\,\big(w(E)\big)>0$
in (\ref{9}). As will be seen in the simplest case of one channel
with lowest threshold, the $2^3P_1$ pole $E_n$ occurs in the
attractive zone of the $DD^*$ channel and hence moves down with
increasing coupling.

\begin{table}[t]
\caption{Charmonium spectrum in the single-channel approach
derived from RSH (\ref{H_0})
\cite{Badalian:2008bi,*Badalian:2004gw,*Badalian:2008dv,*Badalian:2007zz}.
The experimental numbers are taken from PDG \cite{Yao:2006px}. All
masses are in GeV. \label{tab.Mass}}
\begin{center}
\begin{tabular}{ccc}
  \hline\hline
  State (Thresholds) & Theory  & Experiment \\
  \hline
  $1S$ & 3.068 & 3.068 \\
  $1P$ & 3.488 & 3.525 \\
  $2S$ & 3.678 & 3.674 \\
  $(D\bar{D})$ &  & 3.729 \\
 % \multicolumn{3}{c}{$M_{th}(D\bar{D})=3.729$}\\
  $1D$ & 3.787 & 3.771 \\
  $(D^*\bar{D})$ &  & 3.872 \\
 % \multicolumn{3}{c}{$M_{th}(D^*\bar{D})=3.872$}\\
  $2P$ & 3.954 & 3.930 \\
  $(D^*\bar{D^*})$ &  & 4.014 \\
 % \multicolumn{3}{c}{$M_{th}(D^*\bar{D^*})=4.014$}\\
  $3S$ & 4.116 & 4.040 \\
  \hline\hline
\end{tabular}
\end{center}
\end{table}

\begin{figure}[t]
\begin{minipage}[h]{0.45\linewidth}
  \center{\includegraphics[angle=270,width=1.0\textwidth]{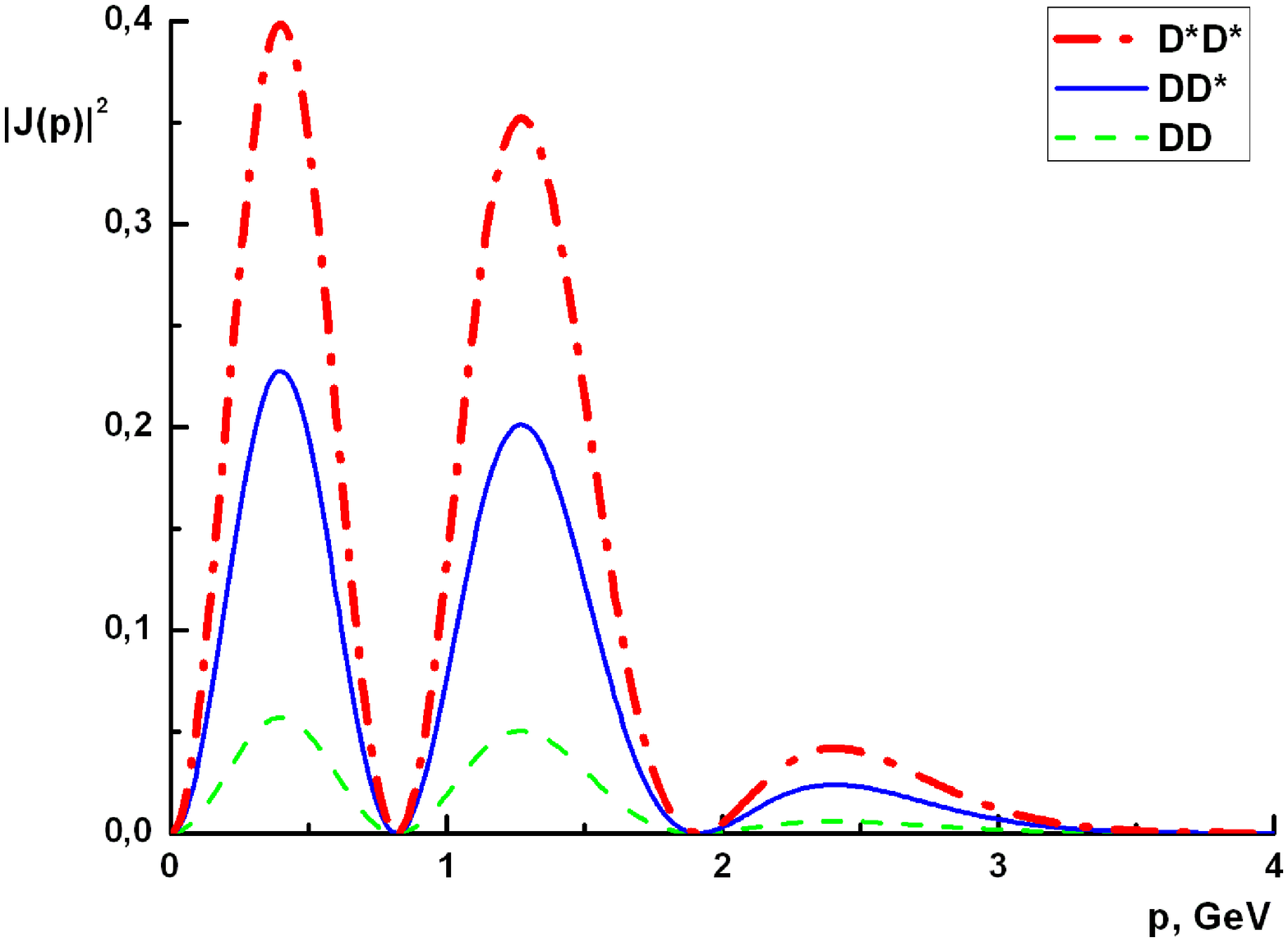}}
  \caption{The squared overlap integral $\frac{1}{3}\sum\limits_{ijk} |J_{n_1n_2n_3}(p)|^2$ for $3^3S_1$
  state.
% and $D\bar{D}$, $D^*\bar{D}$, $D^*\bar{D^*}$ thresholds
  }\label{fig.J(p)}
  \end{minipage}
\begin{minipage}[h]{0.45\linewidth}
\center{\includegraphics[angle=270,width=1.0\textwidth]{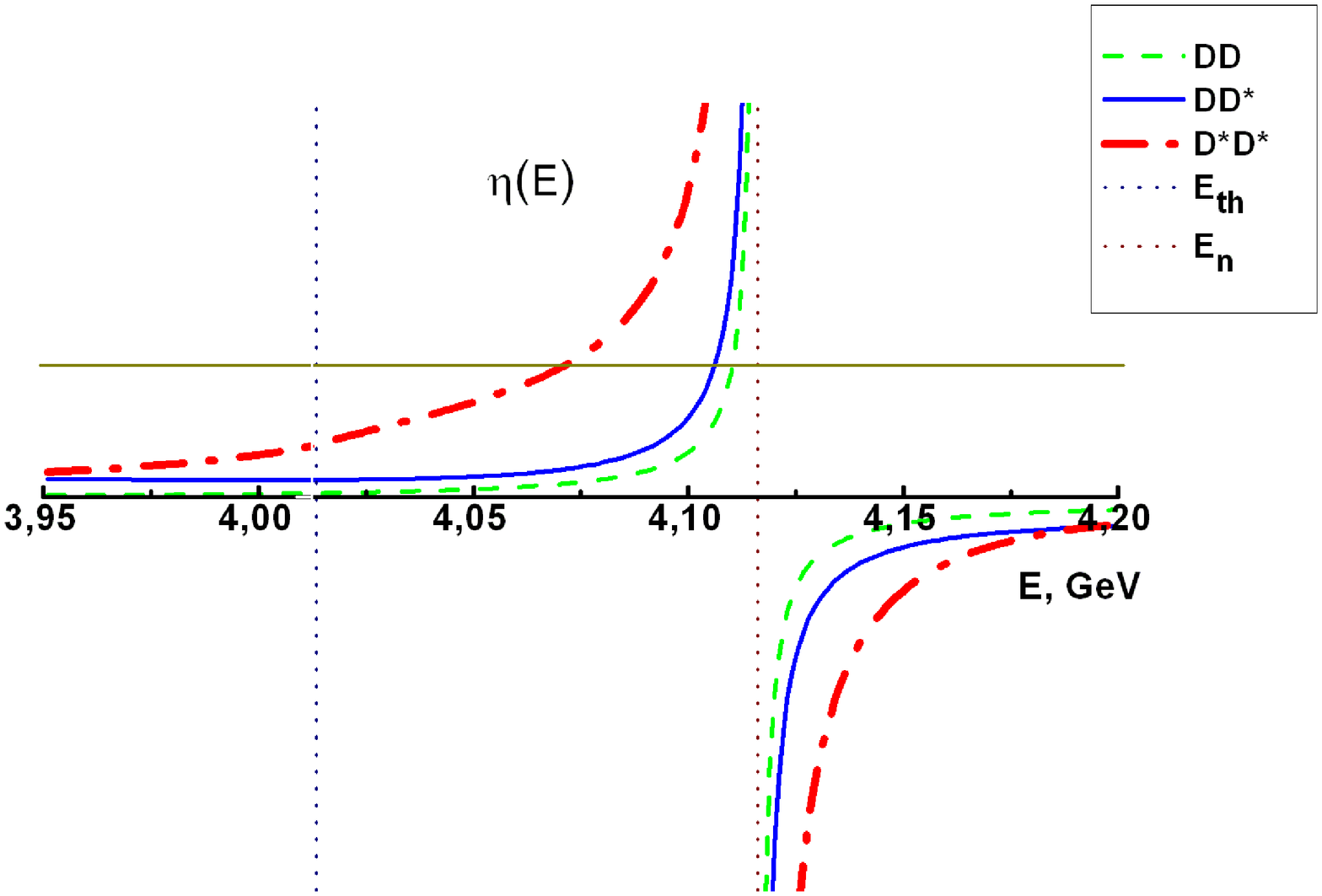}}
\caption{The Weinberg eigenvalue $\textrm{Re}\,(\eta(E))$ for
$3^3S_1$ state, where
 $E_n$=4.116 GeV is the eigenvalue of the single-channel relativistic string Hamiltonian $H_0$ (bare
 resonance) and $E_{th}(D^*\bar{D^*})$=4.014 GeV denotes the closest threshold.}\label{fig.3S_eta}
\end{minipage}
\end{figure}
Below we give several examples of WEM application to different
problems in CC dynamics. We shall consider

i) How CC interaction changes $n^3S_1$ states as compared to
one-channel calculations. We will calculate  energy shifts and
widths for $3^3S_1$ state and also mixing between $3^3S_1$ and
$2^3S_1$ states.

ii) We calculate eigenvalues and amplitudes in the $1^{++}$ state
in connection with the bare $2^3P_1$ level and resulting $X(3872)$
resonance.

To illustrate this formalism we will consider situation with one
level in sector I and one (or many) level(s) in sector II. In
Table~\ref{tab.Mass} we present charmonium mass spectrum in the
single-channel approach (SCA) derived from RSH (\ref{H_0}) (see
for example
\cite{Badalian:2008bi,*Badalian:2004gw,*Badalian:2008dv,*Badalian:2007zz})
in comparison with experimental data and showing the thresholds.

\subsection{\emph{$^3S_1$ levels}}

As a first numerical example we consider the mass shifts and widths of the
$n^3S_1$ $(n=1,2,3)$ states. For these levels the corresponding
$\bar{y}^{red}_{123}$ factors are $\bar{y}^{red}_{123}(^3S_1\rightarrow
D\bar{D})=\frac{q_i}{\sqrt{2}}$; $\bar{y}^{red}_{123}(^3S_1\rightarrow
D^*\bar{D})=i\,\epsilon_{ijm}\,q_m$; $\bar{y}^{red}_{123}(^3S_1\rightarrow
D^*\bar{D^*})=\frac{1}{\sqrt{2}}\,\big(\delta_{ij}\,q_k-\delta_{jk}\,q_i+\delta_{ik}\,
q_j\big)$ (see Appendix B,C) and the transition matrix element is rewritten in
the following form  (see appendix C and equation (\ref{C2a}))
\begin{eqnarray}\label{J(p)}
    J_{n_1n_2n_3}(\textbf{p})&=&\frac{\gamma}{\sqrt{
    N_c}}\,\int\frac{d^{\,3}
    \veq}{(2\pi)^3}\, \bar{y}^{red}_{123}(\textbf{p},\,\textbf{q})\,
    \Psi^{(n_1)}_{Q\bar{Q}}(c\,\textbf{p}+\textbf{q})\,\psi^{(n_2)}_{Q\bar{q}}(\textbf{q})\,\psi^{(n_3)}_{\bar{Q}q}(\textbf{q}),
\end{eqnarray}
where $\gamma\approx1.4$ is channel coupling parameter which is
proportional $M_{\omega}$ (see Appendix C) and
$c=\frac{\omega_Q}{\omega_q+\omega_Q}\simeq 0.73$, where the
averaged kinetic energies  of heavy and light quarks in D meson
$\omega_{q}\simeq0.55$ GeV, $\omega_{Q}\simeq1.5$ GeV are taken
from \cite{Badalian:2007km}. In Eq.(\ref{J(p)})
$\Psi^{(n_1)}_{Q\bar{Q}},\,...=R^{(n_1)}_{Q\bar{Q}}/\sqrt{4\pi},...$
are series of oscillator functions, which are fitted to realistic
w.f. (see Appendix A). We obtain the latter  from the solution of
RSH (\ref{H_0})
\cite{Badalian:2008bi,*Badalian:2004gw,*Badalian:2008dv,*Badalian:2007zz}.

The widths  and mass shifts are obtained from
$|J_{n_1n_2n_3}(\vep)|^2$ averaging over  initial (i)  and summing
over  final (k,j) polarizations. Note that the final formulas for
the  width in channels $DD$, $D\bar{D}^*$ and $D^*\bar{D}^*$
differ by spin factors, which  yield the ratio 1:4:7. From
Eq.(\ref{14}, \ref{43}) one can write the width taking into
account relativistic corrections
\begin{equation}\label{}
\Gamma_{n_1n_2n_3}(\vep)=\frac{p}{\pi}\,|J_{n_1n_2n_3}
(\vep)|^{\,2}\,\left(\frac{1}{\sqrt{\vep^2+M_{n_2}^2}}+\frac{1}{\sqrt{\vep^2+M_{n_3}^2}}\right)^{-1},
\end{equation}
where $M_{n_2},~M_{n_3}$ are the masses of the  corresponding $D$
mesons.
\begin{table}[t]
\caption{The decay width $\Gamma_{n_1n_2n_3}(\vep)$ of
$3^{\,3}S_1$ charmonium state.
%(for channel coupling parameter$\gamma= 1.4$ GeV)
The resonance momentum $p$ is taken from PDG
\cite{Yao:2006px}.\label{tab.Width}}
\begin{center}
\begin{tabular}{lll}
\hline\hline
Channel               & p, GeV & $\Gamma(p)$, MeV \\
\hline
$D\bar{D}$            &0.777& 0.31 \\
$D^*\bar{D}$          &0.576& 25.5 \\
$D^*\bar{D}^*$        &0.227& 17.8 \\
\hline\hline
\end{tabular}
\end{center}
\end{table}
\begin{table}[t]
\caption{Ratios of branching fractions for $3^{\,3}S_1$ state.
%(for channel coupling parameter $\gamma= 1.4$ GeV)
\label{tab.Ratio}}
\begin{center}
\begin{tabular}{llll}
\hline\hline
Ratio     & Experiment \cite{Aubert:2009xs}         &   This paper &$^3P_0$ \cite{Barnes:2005pb,*Barnes:1991em}\\
\hline
$\mathcal{B}(\psi(4040)\rightarrow D\bar{D})/\mathcal{B}(\psi(4040)\rightarrow D^*\bar{D})$    & $0.24\pm0.17$       &0.012& 0.003   \\
$\mathcal{B}(\psi(4040)\rightarrow D^*\bar{D^*})/\mathcal{B}(\psi(4040)\rightarrow D^*\bar{D})$& $0.18\pm0.18$       &0.70 &1.0\\
\hline\hline
\end{tabular}
\end{center}
\end{table}
\begin{table}[h]
\caption{Mass shifts (in MeV) of the $n^3S_1$ states with
$n=1,2,3$.
%(for channel coupling parameter $\gamma= 1.4$ GeV)
\label{tab.Shift}}
\begin{center}
\begin{tabular}{ccccc}
  \hline  \hline
%&\multicolumn{4}{c}{ Mass Shifts}\\
  State & $DD$ & $D\bar{D}^*$ & $D^*\bar{D}^*$ & Total \\
  \hline
  $1^3S_1$ & -5 & -19 & -30 & -54 \\
  $2^3S_1$ & -15& -41& -56& -112 \\
  $3^3S_1$ & -6 & -10 & -45 & -61 \\
  \hline  \hline
\end{tabular}
\end{center}
\end{table}
It is important that the value of the decay width strongly depends
on transition matrix element. This is illustrated by the behavior
of $|J_{n_1n_2n_3}(\vep)|^{\,2}$ for $3^{\,3}S_1$ state. As it can
be seen from Figure~\ref{fig.J(p)}, $|J_{n_1n_2n_3}(\vep)|^{\,2}$
is oscillating and has  two zeros, corresponding to the wave
function nodes. In the small width approximation (\ref{14}) the
width and shift of $E_n$ level will vanish when $p\,(E_n)$
approaches zero on Figure \ref{fig.J(p)}. It is not  a physical
situation, and in the next approximation one should solve
Eq.(\ref{3}) in complex plane and take into account possible
mixing between states due to open channels. For instance it can be
3S-2S, or 3S-2D mixing. Due to the mixing, the w.f. of the "pure"~
states changes and minima in Figure \ref{fig.J(p)} can be filled
in by admixed states. In Tables \ref{tab.Width}, \ref{tab.Ratio}
are given the small width values for $3S$ state of charmonium in
the $D\bar D$ channel, illustrating the zeros discussed above.

In WEM the shifted level positions are defined from Eq. (\ref{40})
and for $3^{\,3}S_1$ one obtains the picture shown in Figure
\ref{fig.3S_eta}. The level shifts calculated from Eq.(\ref{14})
are given in Table \ref{tab.Shift}. One can note relatively small
shifts ($\Delta E \la 100$ MeV) as compared to
\cite{Kalashnikova:2005ui,*Baru:2003qq,Barnes:2007xu}, where
$^3P_0$ and SHO model was used, whereas in our case more
complicated realistic wave functions were exploited.

In addition  we have considered mixing between $3^{\,3}S_1$ and
$2^{\,3}S_1$ levels via $D^*\bar{D}^*$ threshold, which turned out
to be  small, with the mixing angle (defined as in (\ref{69}))
 $\varphi=5^\circ$.

\subsection{\emph{$2^3P_1$-level}}

\begin{figure}[t]
\begin{minipage}[h]{0.45\linewidth}
  \center{\includegraphics[angle=270,width=1.0\textwidth]{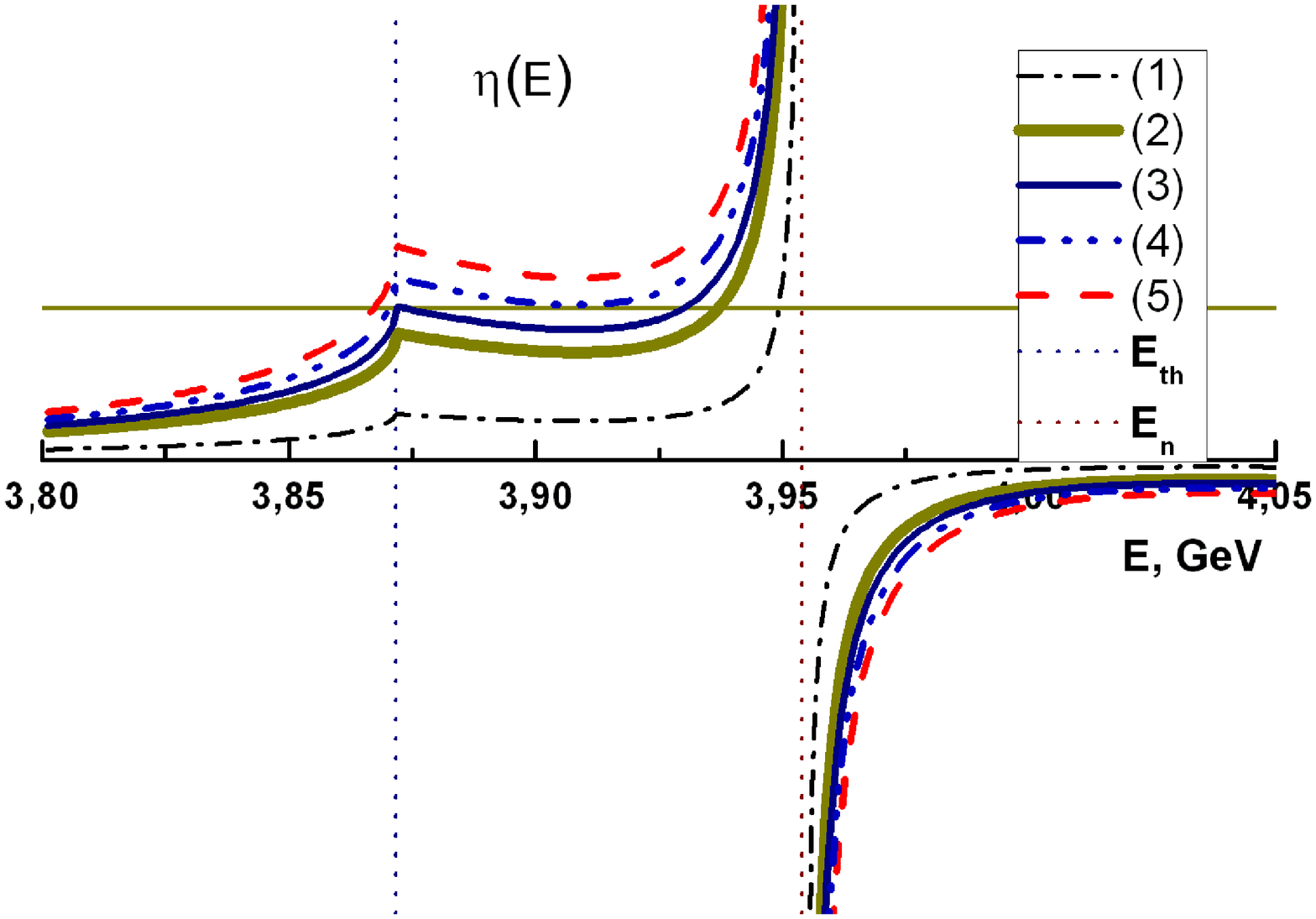}
  (a) One threshold, $E_{th}(D_0D^*_0)$=3.872 GeV.}
  \end{minipage}
\begin{minipage}[h]{0.45\linewidth}
  \center{\includegraphics[angle=270,width=1.0\textwidth]{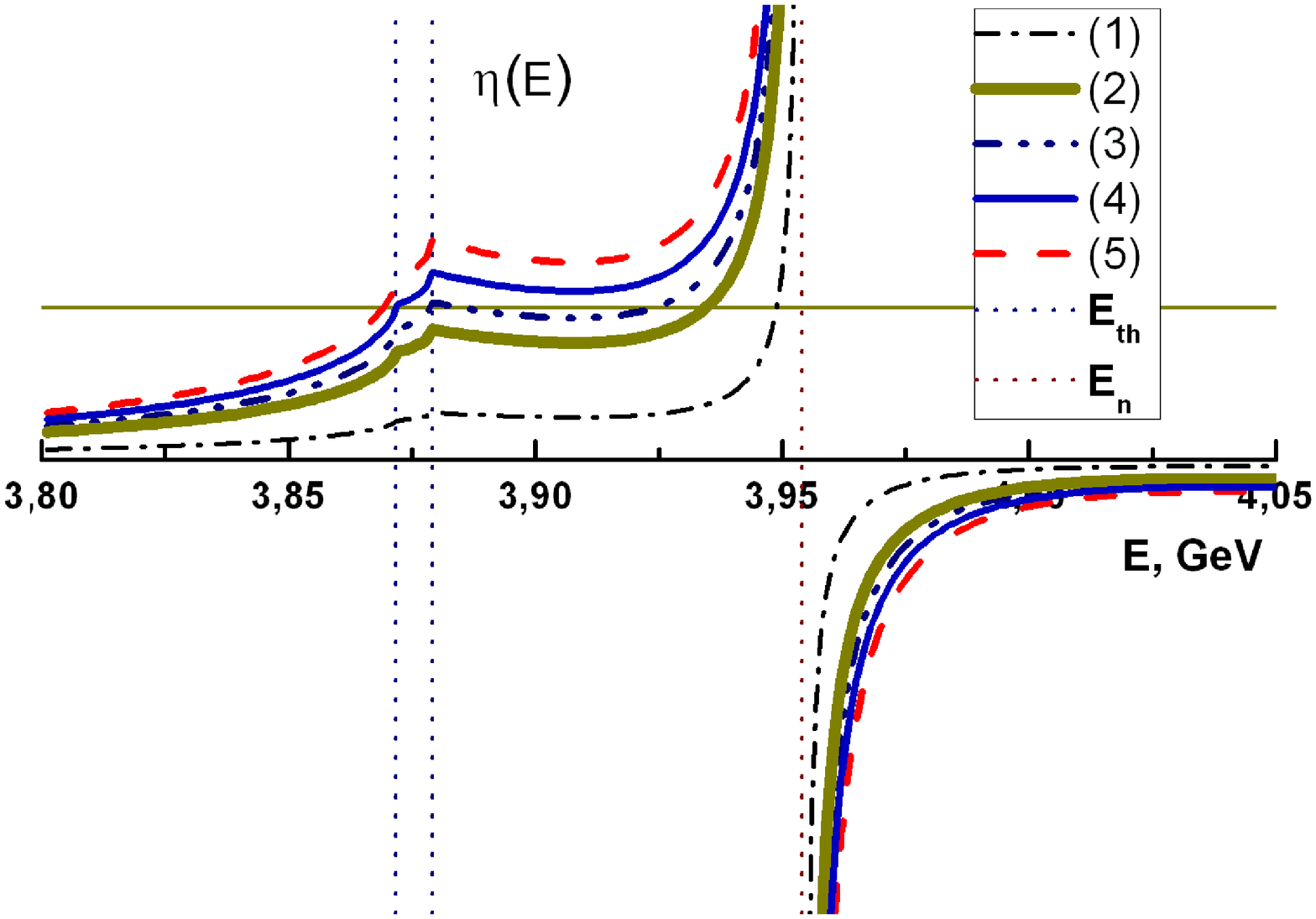}
  (b) Two thresholds, $E_{th}(D_0D^*_0;D_+D_-^*)$=3.872; 3.879 GeV.}
  \end{minipage}
\caption{ The Weinberg eigenvalue $\textrm{Re}\,(\eta(E))$ for
$2^{\,3}P_1$ state with different values of channel coupling
parameter ((1) - $\gamma=0.6$, (2) - $\gamma=1.0$, (3) -
$\gamma=1.1$, (4) - $\gamma=1.2$, (5) - $\gamma=1.3$), $E_n$=3.954
GeV is the eigenvalue of the single-channel relativistic string
Hamiltonian $H_0$ (bare resonance) and
$E_{th}(D_0D^*_0;D_+D_-^*)$=3.872; 3.879 GeV denote
thresholds.}\label{fig.2P_eta}
\end{figure}

\begin{figure}[t]
\begin{center}
  \includegraphics[angle=270,width=0.8\textwidth]{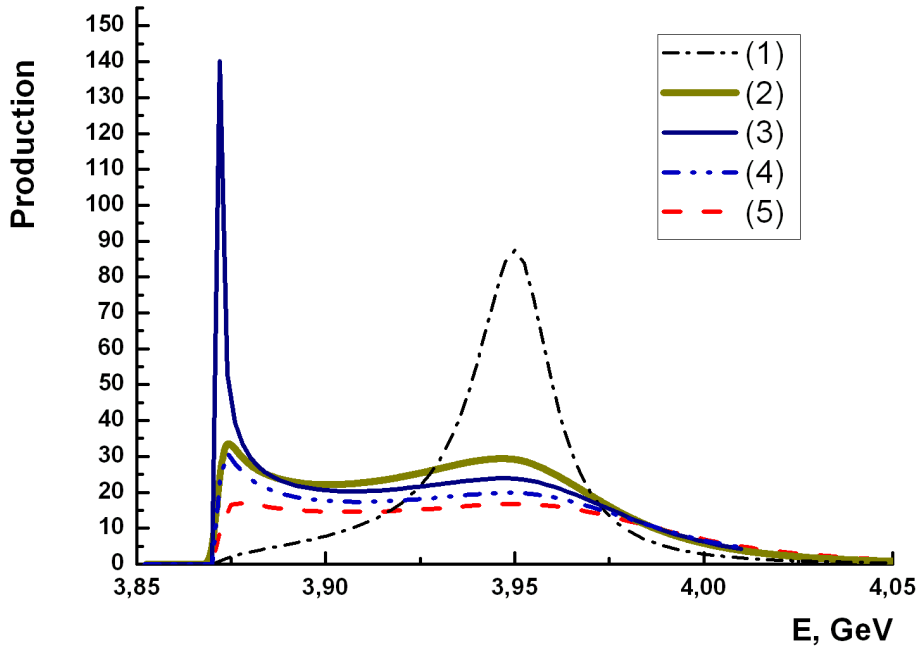}
  \caption{Production crossection  $\sim\frac{\textrm{Im}\,w_{nn}(E)}{|E-E_n-w_{nn} (E)|^2}$ for
$1^{++}$ state with different values of channel coupling parameter
((1) - $\gamma=0.6$, (2) - $\gamma=1.0$, (3) - $\gamma=1.1$, (4) -
$\gamma=1.2$, (5) - $\gamma=1.3$).  For small values of channel
coupling parameter $\gamma$ (line (1)) one can see a good
Breit-Wigner shape, which corresponds to the shifted $2^{\,3}P_1$
state, while for larger $\gamma$ (lines (2), (3), (4)) there is a
broadening of higher resonance together with steep rise near the
threshold $E_{th}(D\bar{D^*})=3.872$ GeV.}
  \label{Production}
\end{center}
\end{figure}

A separate discussion is  needed for the  s-wave decay to charmed
mesons.  We take as an explicit example   the decay
$2^{\,3}P_1\rightarrow DD^*$. Note, that due to positive  C
parity   the s-wave strength is  mostly concentrated in the $DD^*$
channel. In this case, the situation of Figure \ref{Fig.2}(c,d) is
realized when $\textrm{Re}\,\big(\eta(E)\big)$ can cross the unity
line at several energy values, thus producing several resonances.
In our calculations we show $\textrm{Re}\,\big(\eta(E)\big)$ in
Figure~\ref{fig.2P_eta} which correspond to different values of
channel coupling parameter in the region $\pm \,30\%$ around the
standard value $\gamma=1.4$ ($M_\omega=0.8$ GeV). As it can be
seen, $\textrm{Re}\,\big(\eta(E)\big)$ intercepts the line
$\textrm{Re}\,\big(\eta(E)\big)=1$ three times. However we have to
take into account imaginary parts above the thresholds. The
simplest way is to calculate factor
$\frac{|\eta(E)|^2}{|1-\eta(E)|^2}$ which appears in the  squared
t-matrix (\ref{36}). The result is the  two-resonance structure,
one of which is near threshold $M\sim 3.872$ GeV  and another one
near $M\sim3.940$ GeV, the latter  becomes increasingly broad with
increasing coupling $\gamma$ to open channel. In the recent  work
\cite{Kalashnikova:2009gt} a similar form of the first peak was
suggested.
We note, that the factor $\frac{|\eta(E)|^2}{|1-\eta(E)|^2}$ is
relevant for the t-matrix of $DD^*$ scattering, while new
charmonium resonances were observed in production cross sections
like $e^+e^-\to D D^*$ or $B\to K(DD^*)$. Therefore we define the
production yield $|A_3(E)|^2, $ given in (\ref{71}) and show in
Figure \ref{Production} the quantity
$\frac{\textrm{Im}\,w_{nn}(E)}{|E-E_n-w_{nn} (E)|^2}\sim
\frac{\textrm{Im}\, \eta (E)}{|1-\eta(E)|^2\,(E-E_n)}$. In our
approximation ($D_0D^*_0$ and $D_+D_-^*$ thresholds coincide and
there is no connection to $\omega J/\psi$ and $J/\psi\,\pi\pi$
channels) one can see the double peak structure for $\gamma=1.0$;
the first peak at $3.872$ GeV is accompanied by a wide peak around
3.940 GeV.
However, with increasing $\gamma$, when $\gamma=1,2$, the peak in
Fig. \ref{Production} at  3940  becomes flat, while the lower peak
at 3.872 GeV is narrow and high. This picture corresponds to the
experimental situation.

\begin{figure}[t]
\begin{center}
  \includegraphics[angle=0,width=0.8\textwidth]{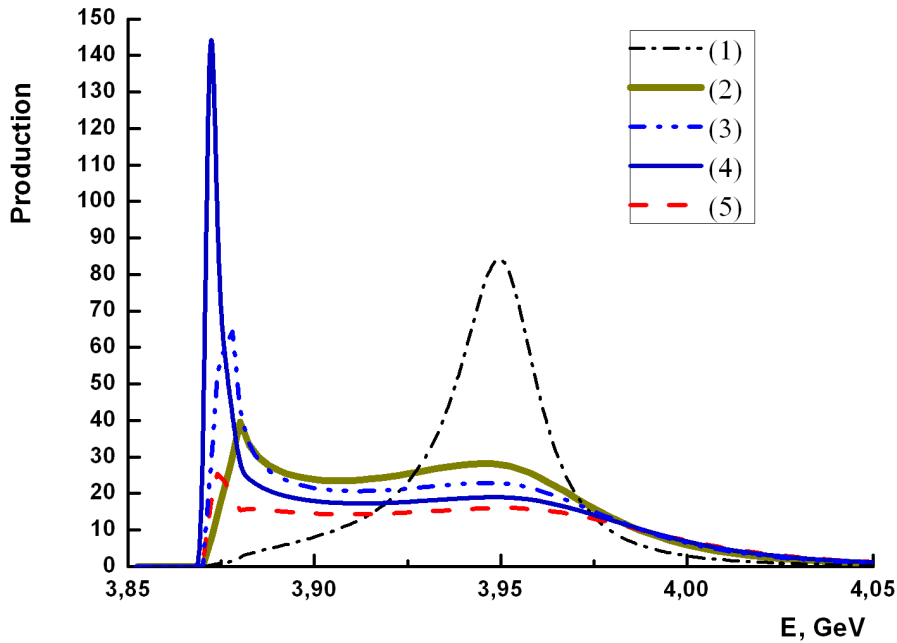}
  \caption{Production cross-section  $\sim\frac{\textrm{Im}\,w_{nn}(E)}{|E-E_n-w_{nn} (E)|^2}$ for
  $1^{++}$ state with different values of channel coupling parameter ((1) - $\gamma=0.6$, (2) - $\gamma=1.0$, (3) - $\gamma=1.1$, (4) -
$\gamma=1.2$, (5) - $\gamma=1.3$) in case, when both thresholds
  $D_0D^*_0$ and $D_+D_-^*$ are taken into account separately.  For small values of channel coupling parameter $\gamma$ (line (1)) one can see a good Breit-Wigner shape,
   which corresponds to the shifted $2^{\,3}P_1$ state, while for larger $\gamma$ (line (2)) there is a broadening together with
   a cusp first near the closest threshold $E_{th}(D_+D_-^*)=3.879$ GeV and then  for $\gamma =1.2$ (line (4)) a sharp peak appears at
   the $E_{th}(D_0D^*_0)=3.872$ GeV.}
  \label{Production_2P_two}
\end{center}
\end{figure}

The case when both thresholds $D_0D^*_0$ and $D_+D_-^*$ are taken
into account, is illustrated by Figure \ref{fig.2P_eta} for
$\textrm{Re}\,\big(\eta(E)\big)$ and Figure
\ref{Production_2P_two} for the production cross section. As can
be seen, the curves for production cross section depend strongly
on channel coupling parameter $\gamma$. For $\gamma =1.0$ (line
(2)) which is 30\% smaller than the nominal value $\gamma=1.4$
$(M_\omega =0.8$ GeV) one can see a peak at the higher threshold
$D_+D_-^*$ and a wider peak at $3.940$ GeV, however for
$\gamma=1.2$ (line (4)), the $3.940$ GeV peak flattens and
simultaneously(!), the peak appears at the lower threshold
$D_0D^*_0$, while only a week cusp is seen at the higher threshold
$D_+D_-^*$. Surprisingly, the isotopically equivalent thresholds
(which we take into account with equal weight) due to different
position in energy plane, provide finally the asymmetric picture
observed in experiment
\cite{Abe:2007jn,*Pakhlova:2008di,*Yuan:2009iu}.

\section{SUMMARY}
\label{sect.9}

We have formulated  equations for Green's functions of strongly
coupled sectors, where new resonances can appear due to CC
interaction. We found that the best formalism for the CC induced
energy-dependent interaction is the Weinberg eigenvalue method.
Conditions for the poles and their positions were systematically
studied in case of $P$-wave and $S$-wave channel coupling. In the
first case one finds only displacement of poles, while in the
second new resonances appear, and in  the $^3P_1$ case two peaks
at 3.872 and 3.940 GeV were found with the height depending on the
coupling constant $M_\omega$. Moreover, we have shown in
Fig.\ref{Production_2P_two}, that at one value of $M_\omega$ the
lower peak is at the $D_0D_0^*$ threshold (but not at the
$D_+D_-^*$ threshold) and at the same time the upper peak at 3.940
GeV flattens. This situation corresponds to the experimental data
\cite{Abe:2007jn,*Pakhlova:2008di,*Yuan:2009iu} and supports our
dynamical CC mechanism.

Mixing of $n^{\,3}S_1$ states was formulated in WEM and found to
be small, while  shifts of $3^{\,3}S_1$ are of the order 50-80
MeV, which signals necessity of mass renormalization.

The method developed in the present paper, provides  a rigorous
definition of resonance wave functions and mixings in the case of
strongly coupled channels.

\section*{ACKNOWLEDGMENTS}

The authors are grateful to Yu.S.Kalashnikova for numerous
discussions, suggestions and help,  to A.M.Badalian for useful
discussions and remarks, to A.I.Veselov for suggestions and help
with computer programs,  and to A.E.Kudryavtsev for good
questions.

The financial help of {\it Dynasty Foundation} to I.V.D. and RFFI
grant 09-02-00629a and grant NS-4961.2008.2 is gratefully
acknowledged.

%apsrevM
%IEEEtranM
\bibliographystyle{apsrevM}
\bibliography{listbMY}

\section*{Appendix A \\WAVE FUNCTIONS} \label{sect.A}

\setcounter{equation}{0}
\def\theequation{A.\arabic{equation}}

In Eq.(\ref{J(p)}) $R_{Q\bar{Q}}^{(n_1)}$, $R_{Q\bar{q}}^{(n_2)}$
and $R_{\bar{Q}q}^{(n_3)}$ are   series of oscillator wave
functions, which are fitted to realistic wave functions. We obtain
them from the solution of the Relativistic String Hamiltonian
(\ref{H_0}), described in
\cite{Dubin:1993fk,*Badalian:2008sv,Badalian:2008bi,*Badalian:2004gw,*Badalian:2008dv,*Badalian:2007zz,Badalian:2008ik,*Badalian:2009bu,*Badalian:2004xv}.

In position space the basic SHO radial wave function is given by
%(see e.g. the textbook by Flugge \cite{Flugge1971})
\begin{eqnarray}
&&R_{nl}^{\scriptscriptstyle
SHO}(\beta,\,r)=\beta^{3/2}\sqrt{\frac{2(n-1)!}{\Gamma(n+l+1/2)}}\,(\beta
r)^l\,e^{-\beta^2r^2/2}\,L_{n-1}^{l+1/2}(\beta^2r^2)\\
\nonumber &&\int\limits_0^{\infty}\,\big(R^{\scriptscriptstyle
SHO}_{nl}(\beta,\,r)\big)^2\,r^{\,2}\,dr=1
\end{eqnarray}
where $\beta$ is the SHO wave function parameter, and
$L_{n-1}^{l+1/2}(\beta^2r^2)$ is an associated Laguerre
polynomial. The realistic radial wave function can be represented
as an expansion  in the full set of oscillator radial functions:

\begin{table}[h]
\caption{\label{tab.betta}Effective values $\beta$ (in GeV) and
coefficients $c_k$ of the series of oscillator radial wave
functions $R_{kl}^{\scriptscriptstyle SHO}(\beta,\,r)$ which are
fitted to realistic radial wave functions $R_{nl}(r)$ of
charmonium and D meson.}
\begin{center}
\begin{tabular}{lclllll}
\hline\hline
State  & $\beta$ & \multicolumn{5}{c}{Coefficients $c_k$}\\
\hline
\multicolumn{7}{c}{Charmonium}\\
$ 1S$ &0.70& $c_1=0.97796$& $c_2=0.169169$& $c_3=0.117682$& $c_4=0.019694$& $c_5=0.025113$\\

$ 2S$ &0.53& $c_1=-0.11889$& $c_2=-0.972774$& $c_3=-0.134041$& $c_4= -0.142303$& $c_5=0.000142$\\

$ 3S$ &0.46& $c_1=-0.09354$& $c_2=0.149573$& $c_3=0.958816$& $c_4=0.112102$& $c_5=0.183886$\\

$ 2P$  &0.48& $c_1=-0.06271$& $c_2=0.981834$& $c_3=-0.123392$& $c_4=0.127111$& $c_5=0.000495$\\
\multicolumn{7}{c}{D meson}\\
1S     &0.48& \multicolumn{5}{c}{c=1}\\
\hline\hline
\end{tabular}
\end{center}
\end{table}
%
% FIGURE 1
\begin{figure}[h]
\begin{minipage}[h]{0.45\linewidth}
\center{\includegraphics[angle=270,width=1.15\textwidth]{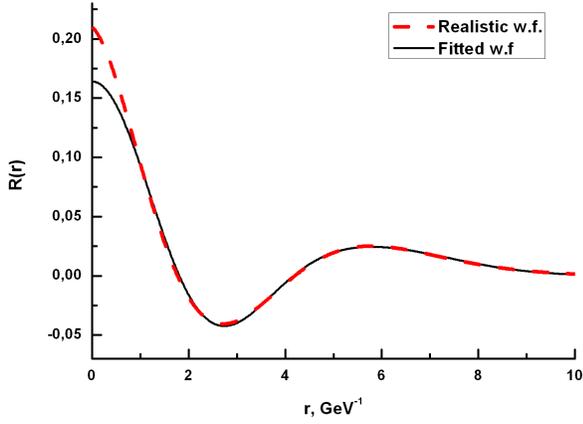} \\
(a) $R^{(n_1)}_{Q\bar{Q}}(3S)/\sqrt{4\pi}$}
\end{minipage}
\hfill
\begin{minipage}[h]{0.45\linewidth}
\center{\includegraphics[angle=270,width=1.15\textwidth]{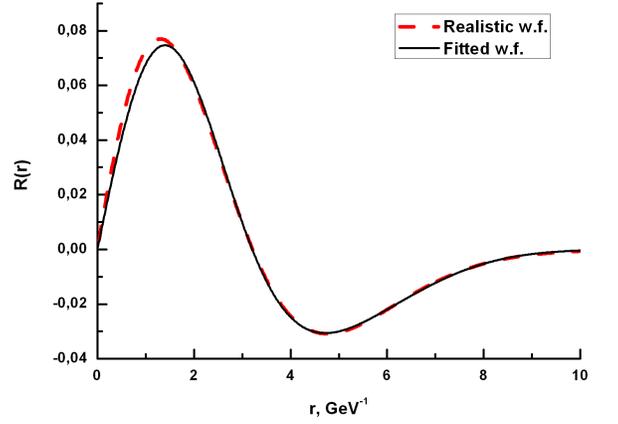} \\
(b) $R^{(n_1)}_{Q\bar{Q}}(2P)/\sqrt{4\pi}$}
\end{minipage}
 \caption{Realistic radial w.f. (divided by $\sqrt{4\pi}$) of charmonium
 $3S$ and $2P$ states (broken lines) and the series of oscillator functions with $k_{max}=5$ (solid lines).
 Note that the solid curves are  almost indistinguishable from the broken ones.}
\label{ris:image1}
\end{figure}
% FIGURE 1
%
\begin{equation}\label{}
    R_{n\,l}(r)=\sum_{k=1}^{k_{max}}c_k\,R_{kl}^{\scriptscriptstyle
SHO}(\beta,\,r).
\end{equation}
Effective values of oscillator parameters $\beta$ and coefficients
$c_k$ are obtained minimizing $\chi^2$ and listed in the Table
\ref{tab.betta}. The quality of approximations can be seen from
the Figure \ref{ris:image1}. In  the momentum space the SHO radial
wave function is given by:
\begin{eqnarray} \nonumber
&&R_{nl}^{\scriptscriptstyle
SHO}(\beta,\,p)=\frac{(-1)^n(2\pi)^{3/2}}{\beta^{3/2}}\,\sqrt{\frac{2(n-1)!}{\Gamma(n+l+1/2)}}\,\left(\frac{p}{\beta}
\right)^l\,e^{-p^2/2\beta^2}\,L_{n-1}^{l+1/2}\left(\frac{p^2}{\beta^2}\right)\\
\nonumber &&\int\limits_0^{\infty}\,\big(R^{\scriptscriptstyle
SHO}_{nl}(\beta,\,p)\big)^2\,\frac{p^{\,2}\,dp}{(2\pi)^3}=1
\end{eqnarray}

\section*{Appendix B \\THE  VERTEX OPERATORS  AND SPINOR  EXPRESSIONS IN THE $(2\times 2)$ FORM} \label{sect.B}
\setcounter{equation}{0}
\def\theequation{B.\arabic{equation}}

Our purpose here is to  go from Eq.(\ref{9}), where $\bar
y^{rel}_{123}$ is the trace of $(4\times 4)$ form to the
$(2\times 2)$ or spinor form, defining in this way  $\bar
y^{red}_{123}.$

We consider operators of the form $(\bar \psi\, \Gamma_i\, \psi)$, with
$\Gamma_i$ consisting of Dirac matrices $\gamma_i$ and derivatives
$\stackrel{\leftrightarrow}{\partial}_i$. To proceed to  the $(2\times 2)$
form, one exploits the limit $M\to \infty$ of the heavy quark mass, so that for
the light quark in the heavy-light meson the Dirac equation can be used, and
one can use symbolically Dirac one-body equation, $\big(\veal \vep + \beta
(m+U)\big)\,\psi =\big(\varepsilon-V\big)\,\psi$, so that for $\psi
=\left(\begin{array}{l}v\\w\end{array}\right)$, one has
$w=\frac{1}{m+U-V+\varepsilon}\, (\vesig\vep)\, v$. One can also use connection
$\bar \psi = C^{-1}\,\psi^c= \psi^c\,(C^{-1})^T$, where $C=(C^{-1})^T=\gamma_2
\gamma_4,~~ \gamma_i =-i\,\beta\, \alpha_i$, so that $ \bar\psi\,\Gamma_i\,\psi
= (v^c, w^c)\, \gamma_2 \gamma_4 \,\Gamma_i
\left(\begin{array}{l}v\\w\end{array}\right)$.
%%%%%%%%%%%%%%%%%%%%%%%%%%%%%%%%%%%%%%%%%%%%%%%%%%%%%%%%%%%%%%%%%%%%%%%%%%%%%%
\begin{table}[h]
\caption{\label{tab.5} Bilinear operators $ \bar \psi\, \Gamma_i\,
\psi$ and their $(2\times 2)$ forms (Notations see in the text).}
\begin{center}
\begin{tabular}{cccc}
\hline \hline
$J^{PC}$& $~^{2S+1}L_J$& $\Gamma_i$ &$ (2\times 2)$ form.\\ \hline
$0^{-+}$& $~^1S_0$& $-i\gamma_5$& $\tilde v^cv-\tilde w^c w$\\

$1^{--}$& $~^3S_1$&$\gamma_ i$& $-(\tilde v^c\sigma_i v+\tilde w^c \sigma_iw)$\\
$1^{+-}$& $~^1P_1$& $-i\gamma_5
\stackrel{\leftrightarrow}{\partial}_i     $& $\tilde
v^c\stackrel{\leftrightarrow}{\partial}_iv-
\tilde w^c\stackrel{\leftrightarrow}{\partial}_i w$\\

$0^{++}$& $~^3P_0$& 1& $i (\tilde v^cw-\tilde w^c v)$\\

$1^{++}$& $~^3P_1$&$\gamma_i\gamma_5$& $-(\tilde v^c\sigma_i w+\tilde w^c \sigma_iv)$\\
$2^{++}$& $~^3P_2$&
$\gamma_i\stackrel{\leftrightarrow}{\partial}_k + \gamma_k
\stackrel{\leftrightarrow}{\partial}_i-\frac23\, \delta_{ik}\hat
\partial$ & $-(\tilde v^c\rho_{ik} v+ \tilde w^c\rho_{ik}
w)$\\

$2^{-+}$& $~^1D_2$& $(\stackrel{\leftrightarrow}{\partial}_i
 \stackrel{\leftrightarrow}{\partial}_k-\frac13\,
\delta_{ik}(\stackrel{\leftrightarrow}{\partial})^2)\gamma_5 $ &
$i (\tilde v^c\omega_{ik}w- \tilde w^c\omega_{ik}
v)$\\
$2^{--}$& $~^3D_2$& $(\gamma_i
\stackrel{\leftrightarrow}{\partial}_k + \gamma_k
\stackrel{\leftrightarrow}{\partial}_i-\frac23\, \delta_{ik}\hat
\partial)\gamma_5$ & $-(\tilde v^c\rho_{ik} w+ \tilde w^c\rho_{ik}
v)$\\ $1^{--}$& $~^3D_1$& $\gamma_i \omega_{ik}$&  $-( v^c\sigma_i
\omega_{ik}v+ \tilde w^c\sigma_i \omega_{ik}
w)$\\
\hline\hline
\end{tabular}
\end{center}
\end{table}
%%%%%%%%%%%%%%%%%%%%%%%%%%%%%%%%%%%%%%%%%%%%%%%%%%%%%%%%%%%%%%%%%%%%%%%%%%%%%%%%%%%%%%%%%%%
Note, that spin indices of charge-conjugated spinors are connected
to ordinary spinors by matrix $\sigma_2$:
$v^c\,\sigma_2=-(\sigma_2\,v^c)^T \equiv \tilde v^c$, and for
$w^c$ one has:
\begin{equation}\label{A2.1}
w^c\,\sigma_2=\left(\frac{1}{m+U-V+\varepsilon}\,\vesig\vep\,
v^c\right)^T \sigma_2=-\tilde v^c \vesig
\overleftarrow{\mathbf{p}}\frac{1}{m+U-V+\varepsilon}\equiv -
\tilde w^c
\end{equation}
where the notation $ \overleftarrow{\mathbf{p}}$ implies, that operator acts on
the left. We are considering 7 lowest states  and display in the Table
\ref{tab.5} the operator $\Gamma_i$,  the corresponding quantum numbers
$J^{PC}$, spectroscopic notation $~^{2S+1}L_J$ and the  equivalent $(2\times
2)$ form for the same vertex $\Gamma_i$ in the last column. We are using in the
Table \ref{tab.5} the following notations
$$ \rho_{ik} \equiv \sigma_i\stackrel{\leftrightarrow}{\partial}_k
+ \sigma_k \stackrel{\leftrightarrow}{\partial}_i -\frac23\, \sigma_l
\stackrel{\leftrightarrow}{\partial}_l \delta_{ik};~~ \omega_{ik} \equiv\,
\stackrel{\leftrightarrow}{\partial}_i\stackrel{\leftrightarrow}{\partial}_k
-\frac13\, \delta_{ik}\,
(\stackrel{\leftrightarrow}{\partial})^2,~~\hat\partial \equiv\,
\stackrel{\leftrightarrow}{\partial}_i\gamma_i.$$ Note, that in the $2\times 2$
form  one has:

\begin{eqnarray}
\nonumber  ^3P_0 &:& \tilde v^c\frac{\vesig
\stackrel{\leftrightarrow}{\vep}}{m+U-V+\varepsilon}\,v \\
\nonumber  ^3P_1 &:&-i e_{ikl} \, \tilde v^c\,
\frac{\stackrel{\leftrightarrow}{p}_k
\sigma_l}{m+U-V+\varepsilon}\, v \\
\nonumber  ^1D_2 &:& \tilde v^c\,\frac{\vesig
\stackrel{\leftrightarrow}{{\vep}}}{m+U-V+\varepsilon}\,\omega_{ik}\,v
\end{eqnarray}

%%%%%%%%%%%%%%%%%%%%%%%%%%%%%%%%%%%%%%%%%%%%%%%%%%%%%%%%%%%%%%%%%%%%%%%%%%%%%%%%%%%%%%%%55
\begin{table}[h]
\caption{\label{tab.6} The $\Gamma^{(n_1)}_{red}$ operator and nonrelativistic
form $\bar y^{red}_{123}$ for $D\bar D$, $D\bar{D}^*$ and $D^*\bar D^*$
channels.}
\begin{center}
\begin{tabular}{cccccc}
\hline\hline
$J^{PC}$& $^{2S+1}L_J$& $\Gamma^{(n_1)}_{red}$&\multicolumn{3}{c}{ $\bar y^{red}_{123}$}\\
\cline{4-6} &&& $D\bar D$& $\frac{1}{\sqrt{2}}\,(D\bar{D}^*\pm \bar{D} D^*)$& $D^*\bar D^*$\\
\hline

$0^{-+}$& $~^1S_0$& $\frac{1}{\sqrt{2}}$
&-&$q_j$&$\frac{i}{\sqrt{2}}\,\epsilon_{jmk}q_m$\\

$1^{--}$& $~^3S_1$&$\frac{1}{\sqrt{2}}\,\sigma_i$&
$\frac{1}{\sqrt{2}}\,q_i$ & $i\,\epsilon_{ijm}q_m$ &
$\frac{1}{\sqrt{2}}\,\big(\delta_{ij}q_k-\delta_{jk}q_i+\delta_{ik}
q_j\big)$\\

$1^{+-}$& $~^1P_1$& $\sqrt{\frac{3}{2}}\,n_i$ &-&$\sqrt{3}\,n_i
q_j$&$ i\,\sqrt{\frac{3}{2}}\,\epsilon_{jmk}q_m n_i $\\

$0^{++}$& $~^3P_0$&  $
\frac{1}{\sqrt{2}}\,\vesig\ven$&$\frac{1}{\sqrt{2}}(\mathbf{q}\mathbf{n})$&-
&$\frac{1}{\sqrt{2}}\,\big(q_kn_j+q_jn_k-\delta_{jk}(\mathbf{q}\mathbf{n})\big)$\\

$1^{++}$& $~^3P_1$&
$\frac{\sqrt{3}}{2}\,\epsilon_{ikl}\sigma_kn_l$ &-&$
i\,\sqrt{\frac{3}{2}}\,\big(q_in_j-(\mathbf{q}\mathbf{n})\delta_{ij}\big)$&-
\\

$2^{++}$& $~^3P_2$& $\frac{3}{4}\,\big(\sigma_i n_l + \sigma_l
n_i-\frac23\, (\vesig\ven) \delta_{il}\big)$
&-&-&$\frac{3}{4}\,\big(n_l(q_k\delta
_{ij}-q_i\delta_{jk}+q_j\delta_{ik})$\\

&&&&&$+n_i(q_k\delta _{lj}-q_l\delta
_{jk}+q_j\delta_{lk})$\\
&&&&&$-n_j(\frac{2}{3}\,q_k\delta_{il})-n_k(\frac{2}{3}\,q_j\delta_{il})$\\

&&&&&$+
\frac{2}{3}(\mathbf{q}\mathbf{n})\delta_{il}\delta_{j,k}\big)$\\
\hline\hline

%$2^{-+}$& $~^1D_2$&$n_in_k -\frac13 \delta_{ik}$&-&&\\ \hline
%
%$2^{--}$& $~^3D_2$& $(\vesig\times \ven)_i n_k$&-&&-\\
%
%&&$+(\vesig\times \ven)_k n_i$&&&\\ \hline
%
%$1^{--}$& $~^3D_1$& $\frac{3}{2}(\mathbf{\sigma n})
% n_i-\frac{1}{2}\sigma_i$ &$\frac{3}{2}\,n_i\,(\mathbf{q}\mathbf{n})-\frac{1}{2}\,q_i$&
% $\frac{i}{\sqrt{2}}\, q_l\,(3n_in_m\epsilon_{mjl}-\epsilon_{ijl})$&$\frac{3}{2}n_i(q_kn_j+q_jn_k-
% (\mathbf{q}\mathbf{n})\delta_{jk})$\\
% &&&&&$-\frac{1}{2}(q_k\delta_{ij}-q_i\delta_{jk}+q_j\delta_{ik})$\\ \hline
\end{tabular}
\end{center}
\end{table}
%%%%%%%%%%%%%%%%%%%%%%%%%%%%%%%%%%%%%%%%%%%%%%%%%%%%%%%%%%%%%%%%%%%%%%%%%%%%%%%%%%%%%%%%%%%%%%%%%%%%

In the $(2\times 2)$ form one can write wave function of
charmonium and D-mesons
$(\Psi^{(n_1)}_{Q\bar{Q}},~\psi^{(n_2)}_{Q\bar{q}},~\psi^{(n_3)}_{\bar{Q}q}$)
as $\Psi_{meson}=\,Const\, \varphi_{nl} (r)\,\big(\tilde v^c
\,\Gamma^{(n)}_{red}\, v\big)$ and normalize it as
\begin{equation}\label{} || \Psi_{meson}||^2 =1 =\int |\varphi_{nl}
(r) |^2 r^{\,2} dr\,  {\tr}\Big\{\Gamma^{(n)}_{red} \Gamma_{red}^{(n)+}\Big\}\,
d\Omega.
\end{equation}
Defining $\varphi_{nl} (r)=R_{nl}(r)/\sqrt{4\pi}$, the
normalization condition for angular part takes the form $\int
{\tr}\Big\{\Gamma^{(n)}_{red}\,\Gamma_{red}^{(n)+}\Big\}\,
\frac{d\Omega}{4\pi} =1$. Then $J_{n_1n_2n_3}(\vep)$ can be
written as in (\ref{8}), but $\bar y^{red}_{123}$ can be found in
spinor $(2\times 2)$ form as
\begin{equation}\label{} \bar y^{red}_{123}= {\tr}\Big\{\Gamma^{(n_1)}_{red}\, \Gamma_{red}^{(n_2)}\, (\vesig
\veq)\, \Gamma_{red}^{(n_3)}\Big\}
\end{equation}
see Table \ref{tab.6}, and all $\Gamma^{(n)}_{red}$ are normalized as written
above. Hence e.g.
$$ \Gamma^{(n_{2,3})}_{red} (D) =\frac{1}{\sqrt{2}},~~ \Gamma^{(n_{2,3})}_{red}
(D^*) = \frac{\sigma_i}{\sqrt{2}}.$$

\section*{Appendix C \\THE PAIR-CREATION VERTEX} \label{sect.C}
\setcounter{equation}{0}
\def\theequation{C.\arabic{equation}}

In the same way  we  consider  here the $(2 \times 2)$ reduction
of the pair-creation vertex, taking $\psi, \bar \psi$ for light
quarks as solutions of Dirac equation and  writing the effective
string-breaking Lagrangian as
\begin{equation}\label{3.9}
\mathcal{L}_{sb} = \int \bar \psi (u)\,   M_\omega\, \psi (u)\,
d^{\,4} u= M_\omega \int \, \frac{i \tilde
v_c\vesig\stackrel{\leftrightarrow}{p} v}{m+U-V+\varepsilon_0}\,
d^{\,4}u
\end{equation}
and we have denoted: $U\equiv \sigma r;$ $V\equiv-\frac43 \frac{\alpha_s}{r};$
$\stackrel{\leftrightarrow}{p} = \vep- \stackrel{\leftarrow}{p}$;
$\varepsilon_0$ is the Dirac eigenvalue $\varepsilon_0 = M_0 (\bar Q q)$
$-M_{\bar Q}$, $\tilde v_c = v_c\, \sigma_2$ is the spinor of  antiquark, $
M_\omega $  is the same as in  Eq. (\ref{3}).

One can take in (\ref{3.9}) the  averaged value of the denominator
$\lan m+U-V+\varepsilon_0\ran\rightarrow m+\lan U\ran - \lan
V\ran+\varepsilon_0$ which effectively redefines our vertex
constant $M_\omega$. As a result the reduced form of the matrix
element $J(p)$  in Eq. (\ref{9}) assumes the form
\begin{equation}
J(\vep) = \frac{\gamma}{\sqrt{N_c}} \int
\frac{d^{\,3}q}{(2\pi)^3}\, \bar y^{red}_{123} (\vep, \veq)\,
\Psi^{+(n_1)}_{Q\bar Q} (c\vep+\veq)\, \Psi_{Q\bar q}^{(n_2)}
(\veq)\, \Psi^{(n_3)}_{\bar Q q} (\veq) \label{C2a}
\end{equation}
where $\bar y^{red}_{123}$ is given in Table \ref{tab.6} and
$\gamma \equiv \frac{2M_\omega}{m+\lan U\ran - \lan
V\ran+\varepsilon_0}$. One can find values of $\lan U\ran$, $\lan
V\ran$, $\varepsilon_0$ in Table \ref{tab.7}, and persuade
oneself, that $\gamma$ is rather stable for different $\alpha_s$
and numerically $\gamma=\frac{2\cdot 0.8 ~{\rm GeV}}{1.2~{\rm
GeV}} \approx 1.4.$

To check consistency of our approximation of putting average
values into denominator, we have compared normlization conditions
of bispinors $(v^+v) + (w^+w)=1=v^+ \left( 1+
\frac{\vep^2}{(\varepsilon + \lan U\ran - \lan
V\ran+m)^2}\right)v$, and found that the term with denominator
contributes around 20\%, and we expect the same accuracy in
definition of $\gamma$. Actually, we are always varying $\gamma$
in the region $\pm 30\%$ around the nominal value $\gamma =1.4$.

One can also check at this point how the $(4\times 4)$ vertex
$M_\omega\, \bar y_{123}$ goes over into the reduced form
$\gamma\, \bar y^{red}_{123}$. E.g. for the $1^{--}$ state
decaying into $DD^*$ one has in the heavy quark mass limit (see
e.g. [13,14]). $\bar y_{123} = \frac{iq_n e_{ikn}}{\omega_q}$, and
$\omega_q \approx 0.6$ GeV is the average energy of the light
quark, which coincides with 1/2 of the denominator in $\gamma$,
while $\bar y^{red}_{123}$ from Table \ref{tab.6} is
$iq_ne_{ikn}$. Thus indeed one has equality $ M_\omega\, \bar
y_{123} = \gamma\, \bar y^{red}_{123}$.

For practical reasons we have used for our calculations the
reduced $(2\times 2)$ forms everywhere.

%%%%%%%%%%%%%%%%%%%%%%%%%%%%%%%%%%%%%%%%%%%%%%%%%%%%%%%%%%%%%%%%%%%%%%%%%
\begin{table}[h] \caption{\label{tab.7} Dirac eigenvalues
$\varepsilon_0$ (in GeV) for quarks of different masses $m$ (in
GeV) and $\alpha_s$.
%The numbers are taken from [25].
The averaged potentials $\lan U\ran$, $\lan V\ran$ (in GeV)  for
different $\alpha_s$ are also presented.} \begin{center}
\begin{tabular}{llll} \hline\hline
$\alpha_s$& 0&0.3&0.39\\
\hline

$m=0.005$ &0.65& 0.493&0.424\\
$m=0.15$&0.80&0.584&0.509\\
$m=0.2$& 0.838& 0.617& 0.539\\
\hline $\lan U \ran$&0.573&0.486&0.463\\
$\lan V \ran$&0&-0.198&-0.273\\ \hline\hline
\end{tabular}
\end{center}
\end{table}
%%%%%%%%%%%%%%%%%%%%%%%%%%%%%%%%%%%%%%%%%%%%%%%%%%%%%%%%%%%%%%%%%%%%%%%%%

In the nonrelativistic limit one has
\begin{equation}
\mathcal{L}_{sb} = i M_\omega \int \tilde v_c \,\frac{\sigma
\stackrel{\leftrightarrow}{p}}{2m}\, v\,d^{\,4}u \label{3.11}
\end{equation}
and for the plane-wave (free) quarks,
$v=\frac{e^{i\,\vek\veu}\,u(\alpha)}{\sqrt{ 2\varepsilon_0 V_3}}$,
one has $\tilde v_c =\frac{\tilde u_c \,e^{-i\,\vek\veu}}{\sqrt{
2\varepsilon_0 V_3}}$
\begin{equation}
\mathcal{L}_{sb} = i \frac{(\tilde u_c\, \vesig
\stackrel{\leftrightarrow}{p}\,u)}{4m}.
\label{3.12}
\end{equation}

%In Table \ref{tab.7} $\lan U\ran$ and $\lan V\ran$ are calculated
%using relativistic string Hamiltonian in
%\cite{Badalian:2008ik,*Badalian:2009bu}.

\section*{Appendix D \\DERIVATION OF EQ.(\ref{36}) etc.} \label{sect.D}
\setcounter{equation}{0}
\def\theequation{D.\arabic{equation}}

To introduce the Weinberg method it is useful to start from the
well-known Hilbert-Schmidt method in integral equations with
symmetric kernels $K(x,y)$, where $x$, $y$ belong to the $n$ -
dimension space. The eigenvalue equation has the form

\begin{equation}\label{D.1}
\phi_n(x)=l_n \int K(x,y)\, \phi_n(y)\,dy
\end{equation}
The spectral decomposition and the resolvent are
\begin{eqnarray}
\nonumber K(x,y)&=&\sum \frac{\phi_n(x) \phi_n(y)}{l_n}\\
\label{D.2}
\Gamma(x,y;l)&=&\sum\frac{\phi_n(x)\phi_n(y)}{(l_n -l)}
\end{eqnarray}
and the orthonormality conditions:
\begin{equation}\label{D.3}
\int \phi_n \phi_m\, dx=\delta_{mn}
\end{equation}

\begin{equation}\label{D.4}
\int \phi_n K(x,y)\,\phi_m \,dx\,dy=\frac{1}{l_n}\,\delta_{mn}
\end{equation}
In the case discussed in section \ref{sect.4}, one arrives to
Eqs.(\ref{32}-\ref{36}), starting from equation
\begin{equation}\label{D.5}
\Psi=-\frac{1}{H_0-E}\, \hat{V}\, \Psi
\end{equation}
and performs symmetrization, using definitions
$\phi_n=\sqrt{H_0-E}\, \Psi_\nu$,

\begin{equation}\label{D.6}
K=-\frac{1}{\sqrt{H_0-E}} \,\hat{V}\, \frac{1}{\sqrt{H_0-E}},\quad
l_n =\frac{1}{\eta_\nu}.
\end{equation}
Now Eq.(\ref{D.3}) yields (\ref{38}), where $a_\nu$ is defined in
(\ref{37}), Eq.(\ref{D.4}) gives (\ref{34}). Similarly, the Greens
function is connected to the resolvent

\begin{eqnarray}
\nonumber G&=&\frac{1}{H_0 -E +V}=\frac{1}{\sqrt{H_0-E}}\,(1 +
\Gamma)\frac{1}{\sqrt{H_0-E}}\\\label{D.7} &=&\sum_\nu\,\Psi_\nu\,
\frac{1}{(1- \eta_\nu)}\, \Psi_\nu\,.
\end{eqnarray}
Now we turn to the $t$-matrix. One has
\begin{equation} t=\hat V -
\hat V\, G\,\hat V;\quad H=H_0 + \hat V\label{A4.8}
\end{equation}
where $\hat V=V_{121}$ in sector I.
One can rewrite (\ref{A4.8})
$$ t= H_0-E+\sum_\nu (H_0-E)\, \Psi_\nu\,\frac{1}{\eta_\nu-1}\,
\Psi_\nu\, (H_0-E)=$$

$$= H_0-E + \sum_\nu \frac{a_\nu (p, E)
\,a_\nu(p',E)}{\eta_\nu -1}=$$

\begin{equation} = H_0-E -\sum_\nu
\frac{\eta_\nu\, a_\nu\, a_\nu}{1-\eta_\nu} -\sum_\nu a_\nu\,
a_\nu.\label{A4.9}\end{equation}
For the latter sum one writes
$$ \sum_\nu a_\nu\, a_\nu=\sum_{\nu,n,n'} c_{\nu n}\,\big(E_n-E\big)\, \Psi_n(p)\,c_{\nu
n'}\,\big(E_{n'}-E\big)\, \Psi_{n'}(p')=$$
\begin{equation}=\sum_{\nu nn'}
\big(E_n-E\big)\, \bar c^\nu_n\, \bar c^\nu_{n'}\, \Psi_n\,
\Psi_{n'} = \sum_n \big(E_n -E\big)\, \Psi_n\, \Psi_n= \lan p\,|\,
\big(H_0 -E\big)\, |p'\ran, \label{A4.10}
\end{equation}
where (\ref{59}) was used.
Hence finally one gets Eq.(36)
\begin{equation}
t=-\sum_{\nu}\frac{\eta_\nu\, a_\nu(p,E)\,
a_\nu(p',E)}{1-\eta_\nu(E)}\,. \label{A4.11}
\end{equation}

\section*{Appendix E \\ANALYTIC STRUCTURE OF WEINBERG AMPLITUDES AND POLE POSITIONS} \label{sect.E}
\setcounter{equation}{0}
\def\theequation{E.\arabic{equation}}

In this Appenix  we study the analytic structure of production and scattering
amplitudes induced by CC resonances. We consider two types of amplitudes, the
scattering amplitude in  the sector II, e.g. $A(D\bar D^* \to D\bar D^*)$, and
production amplitude of the type $(Q\bar Q) \to (Q\bar q)( \bar Q q)$, which
appears in processes e.g. $e^+ e^-\to D\bar D^*,...$ or $B\to K(Q\bar Q) \to
K(D\bar D^*)$.

In the first case the relevant part of amplitude is given in
(\ref{36}), and can be written as

\begin{equation}\label{70} A_1 (E)
=\frac{\eta_\nu(E)}{1-\eta_\nu(E)},~~ {\rm or}~~ A_2 (E)
=\frac{1-\eta_\nu^*(E^\nu)}{1-\eta_\nu(E)}.
\end{equation}

In the second case one can start from (\ref{D.7}) for ($Q\bar Q)$
Green's function and persuade oneself that neglecting mixing of
states one returns to the expression (\ref{11}). The production
crossection is proportional to the imaginary part of $G_{Q\bar Q}$
on the cut, starting from the threshold of interest (e.g. $D\bar
D^*)$ and can be written as
\begin{equation}\label{71}  |A_3 (E)|^2 =\frac{1}{2i}\, \Delta
G^{(I)}_{Q\bar Q} =\sum_n \Psi^{(n)}_{Q\bar Q} (1)\,
\frac{-\textrm{Im}\,\big(w_{nn}(E)\big)}{|E-E_n - w_{nn}
(E)|^2}\,\Psi^{(n)}_{Q\bar Q}(2).
\end{equation}
One can easily find, that the latter expression is proportional to
$\Psi_\nu(1)\,\frac{\textrm{Im}\,\eta_\nu
(E)}{|1-\eta_\nu(E)|^2}\,\Psi_\nu(2)$, so that  of  crucial
importance is the analytic structure of $\frac{1}{1-\eta_\nu(E)}$.

We consider the case, when only one bare $Q\bar Q$ state $E_n$ is
retained, assuming that other states are far off and mixing of
states, discussed in section \ref{sect.5} is unimportant as
compared to the direct influence of the decay channel. In this
case one can write
\begin{equation} \eta_\nu (E) = \frac{w_{nn}
(E)}{E-E_n}\label{72}\end{equation} and we  write $w_{nn}
(E)\equiv w(E)$

We write $w(E)$ as
\begin{equation}\label{A5.1} w(E) =\int
\frac{d^{\,3}\vep}{(2\pi)^3}\, \frac{(J(\vep))^2}{E-E(\vep)} =
\frac{\bar c}{2} \int^\infty_0\frac{\sqrt{u}\, du\, f(u)}{z-u},
\end{equation}
where $\bar c = \frac{\tilde M}{\pi^2},~E(\vep) = E_{th} +
\frac{\vep^2}{2\tilde M},~z= 2\tilde M\,(E-E_{th}) $ and finally
\begin{equation}\label{A5.2}
f(u) = f(\vep^2) = (J(\vep))^2.
\end{equation}
Since $f(\vep^2)>0 $ for all real $\vep^2$, one has
\begin{equation}\label{A5.3} w(0) =-\frac{\bar c}{2} \int^\infty_0
\frac{du}{\sqrt{u}}\, f(u) <0.
\end{equation}
It is convenient to continue  $f(u)$ analitically in the region
near the real axis\footnote{This  is always possible in our
Gaussian ansatz for wave functions and subsequent Fourier
transform $J(p)$, in more general case one continues the
absorptive part, as it is used in the dispersion relation technic,
via the relation
$\textrm{Abs}\,\left(f(E)\right)=\frac{1}{2i}(f^I(E)-f^{II}(E))$,
where $f^i(E)$ is analytic function defined on the $i$-th Riemann
sheet. In the general case one might encounter potential type
singularities  in complex plane, separated from the positive real
axis} and rewrite (\ref{A5.1}) as a contour integral along the
contour $C$ circumjacent the cut in the $u$-plane

$$ w(z) =\frac{\bar c}{4}\, \int_C \frac{\sqrt{u}\, du\, f(u)}{z-u}.$$

It is clear that the same integral along the contour $C'$ with the
point $z$ inside  $C^1$ does not have singularities on the first
sheet, hence one can represent $w(z)$ as follows (difference of
two integrals is the  residue at the pole $u=z$)
\begin{equation}
w(z) =- \frac{i\pi}{2}\, \bar c\, \sqrt{z}\, f(z)+F(z)
\label{A5.4}
\end{equation} where $F(z)$ is a nonsingular
function which can be Taylor expanded around $z=0$.

In (\ref{A5.4}) the argument of $z$ is chosen in a standard way:
$arg(z) =0$, for $z=|z|+i\delta$, and $arg z = \pi$ for $z<0$.

We turn now to the analytic structure of Weinberg amplitudes,
which using (\ref{36}) we write as
\begin{equation}\label{A5.5}
A(E) \equiv \frac{\eta (E)}{1-\eta(E)} = \frac{2\tilde M\,
w(z)}{z-z_p +i \pi \tilde M\, \bar c\, \sqrt{z}\, f(z) - 2\tilde M
F(z)}
\end{equation}
where we have defined $z_p = 4\tilde M\,(E_p-E_{th})$ and $E_p$ is
the bare position of the $Q\bar Q$ level. The denominator in
(\ref{A5.5}) can be rewritten as
\begin{equation}\label{A5.6}
D(z) \equiv z-z_p + i b\,\sqrt{z}\,f (0) - z_p\,\big (1-\eta
(0)\big)+ n(z)
\end{equation}
where we have used relations:
$$\eta (0) = \frac{w(0)}{E_{th} - E_p}= - \frac{2\tilde M w
(0)}{z_p},$$
since $w(0) = F(0)$, and $\eta(0) =- \frac{2\tilde MF(0)}{z_p}$.
We also defined $b=\pi \tilde M\bar c =\frac{\tilde M^2}{\pi}$,
and
$$ n(z) = i b \sqrt{z} \big(f(z) - f(0)\big) - 2 \tilde M\big(F(z) -
 F(0)\big).$$
Since $n0) =0$, we expect it does not affect strongly the analytic structure of
$D(z)$ near $z=0$, where $n(z)$ can be written as
\begin{equation}\label{A5.7}
\eta(z) = c_1 z + i \,d_1\, z^{3/2} + \mathcal{O}(z^2,\,z^{5/2}).
\end{equation}

The poles of $A(E)$ in the zeroth approximation $(n \equiv 0)$ are easily
found, denoting $\sqrt{z}\equiv k$, one has two poles at $k= k_+,\, k_-$, with
\begin{equation}
k_+ = - \frac{i\,bf(0)}{2}+ \sqrt{- \left(
\frac{bf(0)}{2}\right)^2 + z_p
\big(1-\eta(0)\big)}\label{A5.8}\end{equation}
\begin{equation} k_- = -\frac{i\,bf(0)}{2} - \sqrt{- \left( \frac{bf(0)}{2}\right)^2 + z_p
\big(1-\eta(0)\big)}\label{A5.9}\end{equation}

Here occur two limiting situations, (i) $z_p$ is small (the bare
pole is in the proximity of the threshold), or
\begin{equation}
z_p \,(1-\eta(0))\ll \left(\frac{bf(0)}{2}\right)^2\label{A5.10}
\end{equation}
(ii) $z_p$ is large, (pole $E_p$ far from threshold)
\begin{equation} |z_p\,(1-\eta(0))|\gg \left(
\frac{bf(0)}{2}\right)^2.\label{A5.11}\end{equation}
In case (i) the poles are (neglecting  higher order terms)
\begin{equation} k_+ =-i\,\frac{z_p\big(1-\eta(0)\big)}{bf(0)}\label{A5.12}\end{equation}
\begin{equation} k_-=-i\,bf(0)+i\,\frac{z_p\big(1-\eta(0)\big)}{bf(0)}\,.\label{A5.13}\end{equation}

One can see, that for weak CC interaction, when $\eta(0)<1$, both
poles are on the second sheet (virtual states), while for strong
CC interaction, $\eta(0)>1$, the pole $k_+$ is a bound state,
while $k_-$ is a virtual state.

Now for the case (ii) one can write
\begin{equation} k_\pm =\pm
\sqrt{z_p (1-\eta(z))}\left(1-\frac12\left(
\frac{b\,f(0)}{2}\right)^2 \frac{1}{z_p(1-\eta(0))}\right)-
\frac{i\,bf(0)}{2}\label{A5.14}
\end{equation}
and in the  standard situation, when $z_1\big(1-\eta(0)\big)>0$,
one has a pair of Breit-Wigner poles $E_0\mp \frac{i\,\Gamma}{2}$,
with
\begin{equation} E_0 =E_1 \big(1-\eta(0)\big) -\left(
\frac{bf(0)}{2}\right)^2 \frac{1}{\tilde
M}\label{A5.15}
\end{equation}
\begin{equation}
\Gamma= \frac{p_p\tilde M}{\pi} f(0), ~~ p_p =\sqrt{2\tilde
M\big(E_p-E_{th}\big)} \label{A5.16}
\end{equation}

Note, that (\ref{A5.16}) coincides with (\ref{14}) as it should
be. Using (\ref{A5.5}), (\ref{A5.7}) one can write the following
analytic representation for the Weinberg amplitude in terms of
variable $k\equiv \sqrt{z}$ \begin{equation} A(k) = \frac{2\tilde
M \left(- \frac{i\,\pi\,\bar c}{2}\, kf (k^2)\right) +
F(k^2)}{(k-k_+)(k-k_-)+ c_1\,k^2 + i\,d_1\, k^3
+\mathcal{O}(k^4)}\label{A5.17}\end{equation}

Note, that for the CC poles $(k_+, k_-$ near threshold) the form
of $A(k)$ is far from the Breit-Wigner type and is of the cusp
type, with infinite energy derivative near the pole, which
possibly explains the very narrow peak of $X(3872)$.

Finally, we discuss the case of several thresholds, e.g. in
$X(3872)$ for isospin  zero one has a sum of $D_0 \bar D_0^*+$
h.c. and $D_+\bar D^*_-+$h.c.  terms in $w$, so that in general
case one can write for $n$ thresholds.
\begin{equation}
w(E)=\sum^N_{i=1} \frac{\bar c_i}{2}\, \int^\infty_0
\frac{\sqrt{u}\, du\, f_i (u)}{z_i-u} \label{A5.18}
\end{equation}
where $z_i =2\tilde M_i\,(E-E^{(i)}_{th}) $. One can apply to
$w(E)$ the same procedure  as before to separate out nonanalytic
terms, with the result, that $D(z)$ now has the form
\begin{equation}
D(z) = z-z_p +i\,b_1\, \sqrt{z}\, f_1(0) +i\,\sum^n_{i=2}
b_i\,\sqrt{z-\Delta_i}\, f_i (-\Delta_i) + n(z) \label{A5.19}
\end{equation} where we have kept notations
for $z$ with respect to the lowest threshold, and $\Delta_i =
2\tilde M_i (E^{(i)}_{th}- E^{(1)}_{th})$.

It is important, that for $z< \Delta_i$ the argument of the square
root term is  $ \left( i\frac{\pi}{2}\right)$ leading to some
renormalization of the term $z_p$ for large $\Delta_i$, while for
small $\Delta_i$ the situation is complicated and should be solved
explicitly in the complex plane $z$.

\end{document}